\documentclass[11pt]{article}

\usepackage{amsmath} 
   \usepackage{amssymb} 
	
\begin{document}
	
	 \newtheorem{thm}{Theorem}
	\newtheorem{proof}{Proof}
 \newtheorem{coro}{Corollary}

\title{%
On some exact solutions of heavenly equations in four 
      dimensions}

\author{{\L}. T. St\c{e}pie\'{n} \thanks{The Pedagogical University of Cracow, ul. Podchora\c{}\.{z}ych 2, 30-084 Krak\'{o}w, Poland} \thanks{e-mail: sfstepie@cyf-kr.edu.pl, lukasz.stepien@up.krakow.pl}} 

\date{} 

 \maketitle

\begin{abstract}%
Some new classes of exact solutions (so-called functionally-invariant solutions) of the elliptic and hyperbolic complex Monge-Amp$\grave{e}$re equations and of the second heavenly equation, mixed heavenly equation, asymmetric heavenly  equation, evolution  form of second heavenly equation, general heavenly equation, real general heavenly equation and one of the real sections of general heavenly equation, are found. Besides non-invariance of these found classes of solutions has been investigated. These classes of solutions determine the new classes of metrics without Killing vectors. A criterion of non-invariance of the solutions belonging to found classes, has been also formulated.
\end{abstract}

keywords: first heavenly equation; second heavenly equation; heavenly equations; functionally-invariant solutions; functional-invariant  solutions; Monge-Amp$\grave{e}$re equation

    \section{Introduction} 
   This is well-known fact that Einstein equations of gravitational field do not appear to be generically integrable. In order
   to find exact solutions, one needs to consider spacetimes with symmetries. There in \cite{stephani_etal}, is a survey including 
   exact solutions of Einstein equations. One of such cases of symmetries of spacetime, is this, when the spacetime has
   two Killing vector, then the field equations reduce to so-called Ernst equation, \cite{ernst}. This equation has been discussed in  
   \cite{klein_richter}. 
  Another interesting and well-known case of reducing of Einstein equations, is the case of the so-called, second heavenly equation.    It was derived by Pleba\'{n}ski in \cite{plebanski1975}. The symmetries of this equation were investigated in, \cite{boyerwinternitz}. In \cite{abraham2007} some hidden symmetry of type II and some exact solutions of the second heavenly equation have been obtained, by a 
  reduction of this equation to the homogeneous Monge-Amp$\grave{e}$re equation in similarity 
  variables. One of the very important classes of solutions of elliptic complex Monge-Amp$\grave{e}$re equation, are the
  solutions, which generate metrics, not possessing Killing vector, because such solutions are non-invariant and so they are the candidates for gravitational instantons, \cite{malykh2004}. Gravitational instantons can be defined analogically to instantons
	in Yang-Mills theory, as solutions of Einstein equations, non-singular on a section of complexified spacetime, where the curvature
	decays at large distances, \cite{Hawking1977}. 
	The most wanted gravitational instanton
  is so called, Kummer surface $K3$, \cite{malykh2004}, \cite{malykh2003}, \cite{atyiah_hitchin_singer}. $K3$ with a Ricci-flat Calabi-Yau metric and the complex torus with the flat metric, are the only compact 4-dimensional Riemannian hyper-K\"{a}hler manifolds, \cite{dunajski_ksiazka}. 
Finding of explicit form of the metric corresponding to Kummer surface $K3$ is a challenging problem, among others, because of the requirement
  of non-existence of Killing vector for such metric, which implicates the requirement of non-invariance of the solution of homogeneous elliptic complex Monge-Amp$\grave{e}$re equation. 
	This stated the motivation of our searching for non-invariant solutions of homogeneous elliptic complex Monge-Amp$\grave{e}$re equation, however, as it has turned out, it is hard to find simultaneously non-invariant and real solutions in this case and just such solutions can describe Kummer surface.
  Some such exact solutions have been  obtained in \cite{malykh2003} and there also non-existence of Killing vector for these solutions has been checked. In \cite{malykh2004} the real solutions of the hyperbolic complex Monge-Amp$\grave{e}$re equation was found and their non-invariance was also be checked. In \cite{Robaszewska} was showed that one can describe locally, non-degenerate complex surfaces by a solution for some Monge-Amp$\grave{e}$re type equation. Some exact solutions of multidimensional Monge-Amp$\grave{e}$re equation were found in \cite{Fedorczuk}, by using the subgroup of the generalized  Poincar$\grave{e}$ group $P(1,4)$. In \cite{Kushner} contact transformations were appplied to Monge-Amp$\grave{e}$re type equation. Some aspects of complex Monge-Amp$\grave{e}$re equation (among others, the existence and stability of its weak solutions on compact K\"{a}hler manifolds), were studied in \cite{Kolodziej}. 
	In \cite{KuznetsovaPopowiczToppan}  the $sl(2n|2n)^{1}$ super-Toda-Lattices and the heavenly equations as continuum limit were investigated. The $N=2$ heavenly equation is studied in \cite{toppan}.
	In \cite{wazwaz} the multikink solutions of second heavenly and asymmetric heavenly equations have been obtained. 
  In \cite{sheftel2009} a classification of scalar partial differential equations of second order, non-invariant solutions of mixed heavenly equation and a connection between this equation and Husain equation have been presented. On the other hand, there in \cite{jakim} it has been showed that every solution of Husain equation 
  (related to chiral model of self-dual gravity, \cite{husain}) defines some  solution of well-known Pleba\'{n}ski first heavenly equation.\\       
  There in \cite{sheftel2009} also so-called asymmetric heavenly equation has been derived. This equation is connected to so-called evolution form of second heavenly equation, \cite{Finley_Plebanski_etal}, \cite{Plebanski_Przanowski}, \cite{Boyer_Plebanski}. In \cite{Finley_Plebanski_etal} the evolution form of second heavenly equation has been derived by some symmetry reduction of Lie algebra of the area preserving group of diffeomorphisms of the 2-surface $\Sigma^{2}$ of self-dual Yang-Mills equations. The same result
  has been obtained in the case of second heavenly equation in \cite{Plebanski_etal}. 
In \cite{malykh2011JPA} it was showed that the so-called general heavenly equation governed anti-self-dual (ASD) gravity and also
some exact solutions of this equations were presented. 
  
  In this paper we show that by applying so-called decomposition method,  it is possible to find some new classes of exact non-invariant solutions (so-called functionally invariant solutions) of:
	the elliptic and hyperbolic complex Monge-Amp$\grave{e}$re equations, 
	the second heavenly equation, mixed heavenly equation, asymmetric heavenly equations, evolution form of second heavenly equation, and general heavenly equation (also real  general heavenly equation and one of real sections of general heavenly equation). 
  The main version of the decomposition method,  mentioned above, has been presented in \cite{stepien2010}.\\
  
  This paper is organized as follows. In section 2 we briefly describe the procedure of the decomposition method (of course, one should not confuse this method with the Bogomolny decomposition, which was obtained for some  nonlinear models in field theory,  
	in the papers \cite{sokalski1984}, \cite{skistska2002} and \cite{stepien2015}). Section 3 includes a short  introduction of the heavenly equations mentioned above and to the non-invariance of the solutions of these equations. In section 4 we find, by using the mentioned decomposition method, the classes of exact solutions of the mentioned equations. Next, in this same section, we investigate non-invariance of the solutions belonging to the classes mentioned above.\\ 
  We formulate also a criterion of non-invariance of the solutions belonging to found classes. In section 5 we give some conclusions. 
  The current paper is a new version of the paper \cite{stepien2012}.

      \section{A short description of the decomposition method} 
      
       Now we shortly describe the decomposition method, introduced for the first time in \cite{stepien2010}. \\ 
       Let's assume that we have to solve some nonlinear partial differential equation
       
        \begin{equation}
    F(x^{\mu}, u_{1},...., u_{m}, u_{1,x^{\mu}}, ..., u_{m,x^{\mu}}, u_{1,x^{\mu}, x^{\nu}}, ...) = 0,
    \label{rownanie}
       \end{equation}
       
       where $u_{n,x^{\alpha}}=\frac{\partial u_{n}}{\partial x^{\alpha}} \hspace{0.08 in} etc.$
       
       According to the assumptions of the decomposition method, which was presented
       first time in \cite{stepien2010}, firstly we check, whether it is possible to 
       {\it decompose} the equation on the fragments, characterized by a {\it homogenity} of 
       derivatives. 
          For example, such decomposed investigated equation may be as follows, \cite{stepien2010} :
           
        \begin{equation} 
 G_{1} \cdot [(u_{,x})^{2} + (u_{,y})^{2}] + G_{2} \cdot [u_{,x x} +  u_{,x y}]=0,    \label{przyklad}
    \end{equation}
    
 where $u=u(x,y)$ is some function of class $\mathcal{C}^{2}$ and the terms: $G_{1}, G_{2}$ may depend on $x^{\mu}, u, u_{x^{\mu}}, ...$ 
         and $u \in \mathbb{R}$ or $u \in \mathbb{C}$, in dependence on investigated problem.

       We see that the result of the checking is positive and then, we insert into (\ref{przyklad}),
       the ansatz, \cite{stepien2010}:
         
          \begin{equation} 
     u(x^{\sigma}) = \beta_{1} + f( a_{\mu} x^{\mu} + \beta_{2} , b_{\nu} x^{\nu} + 
    \beta_{3}, c_{\rho} x^{\rho} + \beta_{4} ) .        \label{ans1}  
        \end{equation}

   In the above ansatz we try to keep $f$ as an {\em arbitrary} function (of class 
   $\mathcal{C}^{2}$), so far as it is possible. The class of the solutions given by the ansatz of such kind, is called in the literature, as {\em functionally invariant} solution, \cite{sobolew}. The function $f$ depends on the appropriate arguments, like this one: $a_{\mu} x^{\mu} + \beta_{2}$. In this paper: $a_{\mu} x^{\mu}=a_{1}x^{1} + a_{2} x^{2} + a_{3} x^{3} + a_{4} x^{4}$. The coefficients $a_{\mu}, b_{\nu}, c_{\rho}$ may be in general complex numbers, which are to be determined later, $\beta_{j}$ may be in general complex constants, $j=1, ..., 4$, and $\mu, \nu, \rho = 1,...,4$. In general, the set of values of $\mu, \nu, \rho$,  depends on the investigated equation. We can decrease or increase the number of the arguments of the function $f$ in (\ref{ans1}) and also modify the form of the ansatz (\ref{ans1}), in dependency on the situation.\\ 
   
   We make such modification later in this section and in the section 4. \\

  After inserting the ansatz (\ref{ans1}) into the example equation (\ref{przyklad}), there, instead of partial derivatives of $u$, the derivatives of the  function $f$ appear: $D_{1} f, D_{2} f, ..., D_{1,1} f, D_{1,2} f, ...$, where the indices denote differentiating with respect to first and so far, arguments of the function $f$ (like this one: $a_{\mu} x^{\mu} + \beta_{2})$.
  
  For example, if we insert a two-dimensional version of ansatz (\ref{ans1}) into (\ref{przyklad}) and collect appropriate terms by the derivatives $D_{j}f$ and $D_{j,k}f$, then, we get, \cite{stepien2010}:
    
		      \begin{eqnarray}
         \begin{gathered} 
    G_{1} \cdot [(a_{1}^{2} + a_{2}^{2}) (D_{1} f)^{2} + (b_{1}^{2} + b_{2}^{2}) (D_{2} 
    f)^{2} + \\ 
		(c_{1}^{2} + c_{2}^{2}) (D_{3} f)^{2} + 
     2(a_{1} b_{1} + a_{2} b_{2}) D_{1} f D_{2} f +  \\
 2(a_{1} c_{1} + a_{2} c_{2} ) D_{1} f D_{3} f+ 2(b_{1} c_{1} + b_{2} c_{2}) D_{2} f 
 D_{3} f] + \nonumber \\ 
 G_{2} \cdot [(a_{1}^{2} + a_{1} a_{2} ) D_{1,1} f + ( 2a_{1} b_{1} + a_{1} b_{2} + 
 a_{2} b_{1} ) D_{1,2} f \\ 
 + ( 2a_{1} c_{1} + a_{1} c_{2} + a_{2} c_{1}) D_{1,3} f +  
 (b_{1}^{2} + b_{1} b_{2} ) D_{2,2} f +  \nonumber \\
 ( 2 b_{1} c_{1} + b_{1} c_{2} + b_{2} c_{1} ) D_{2,3} f + (c_{1}^{2} + c_{1} c_{2}) 
 D_{3,3} f] = 0, \nonumber 
           \end{gathered} 
					 \end{eqnarray} 
   
   where $D_{j}f, D_{j,k}f$ denote correspondingly: a derivative of the function $f$ with respect to
   $j-nary$ argument and the mixed derivative of this function with respect to $j-nary$ and $k-nary$
   argument. \\
           Now, we require that all algebraic terms in the parenthesises must 
         vanish. As a result we obtain a system of algebraic equations, which 
         solutions are the parameters $a_{1}, a_{2},... $. We call such system of 
         algebraic equations as determining algebraic system. Its solutions establish the relations between 
         $a_{\mu}, b_{\nu}, c_{\rho}$ and therefore they constitute, together with (\ref{ans1}), some class of 
         solutions of (\ref{przyklad}). In dependence on the situation, we may need to take into consideration additionally some other conditions, which must be satisfied by our class of solutions. These conditions implicate the requirement of satisfying of some algebraic equations,  
         which we attach to the determining algebraic system. In this paper,  one  example of such additional condition, is the condition of non-invariance of the solutions. 
         The ansatz (\ref{ans1}) appears, as an effect of a generalization of some 
         result, obtained in \cite{sokalski1984} and \cite{skistska2002}. Namely, 
         there in the mentioned papers, 
         some classes of exact solutions of Bogomolny decomposition (Bogomolny equations) for Heisenberg model of ferromagnet have been 
         obtained (but by applying some other method - so called, concept of strong necessary conditions): 
         
         \begin{equation}
   \omega=\omega[(i\alpha + \beta\gamma)x_{1} + (i\gamma - \alpha \beta)x_{2} + (\beta^{2}-1)x_{3}], 
   \hspace{0.1 in} c.c. , \label{ski1984} 
         \end{equation}
         
         where $\omega$ is arbitrary holomorphic function of class $\mathcal{C}^{2}$, depending on its 
         argument and $\omega=\frac{S^{1}+iS^{2}}{1+S^{3}}$, $S^{i}$ ($i=1,2,3$) - components of classical 
         Heisenberg spin, and $\alpha^{2} + \beta^{2} + \gamma^{2}=1$. So, the ansatz (\ref{ans1}) is 
         a generalization of (\ref{ski1984}). 
         The solution (given by the ansatz (\ref{ans1})) has been obtained for: Heisenberg model, 
         nonlinear $\sigma$ model (or O(3) model) and scalar Born-Infeld-like equation in (3+1)-dimensions
         in \cite{stepien2010} and for Skyrme-Faddeev model in \cite{stepien2015}. \\
         
         \vspace{0.5 in}
         
         One can easily show that this method can be extended for the class of the equations of {\em arbitrary} 
         order:
         
				\begin{eqnarray}
        \begin{gathered}
    F(x^{\mu}, u_{1},...., u_{m}, u_{1,x^{\mu}}, ..., u_{m,x^{\mu}}, u_{1,x^{\mu} x^{\nu}}, ..., \\ 
		u_{1, x^{\alpha_{1}}...x^{\alpha_{n}}},...,u_{m,x^{\mu} x^{\nu}}, ..., u_{m, x^{\alpha_{1}}...x^{\alpha_{n}}} ) = 0,
    \label{rownanie_wyzsz}
       \end{gathered}
			 \end{eqnarray} 
       
   obviously, if decomposition on the proper fragments, mentioned above, is possible.\\ 
         
       Of course, this above decomposition method may be also applied for solving linear partial 
          differential equations, homogeneous with respect to the derivatives.\\
          
          Just now, in the order to find classes of non-invariant solutions, we make a modification of 
          (\ref{ans1}) and the basic form of (\ref{ans1}), which be applied for all equations considered
					in this paper, is:

   \begin{gather}
   u(x^{\mu})=\beta_{1} + g_{1}(\Sigma_{1})+g_{2}(\Sigma_{2})+g_{3}(\Sigma_{3})+g_{4}( 
   \Sigma_{4}), \label{rozw_suma} 
             \end{gather}
 
             where:

            \begin{equation}
            \begin{gathered}
             \Sigma_{1} = a_{\mu} x^{\mu} + \beta_{2}, \hspace{0.2 in} \Sigma_{2} = b_{\mu} x^{\mu} + 
             \beta_{3}, \\
                \Sigma_{3} = c_{\mu} x^{\mu} + \beta_{4}, \hspace{0.2 in}
                \Sigma_{4} = d_{\mu} x^{\mu} + \beta_{5}, \label{argumenty1} 
             \end{gathered}
             \end{equation}
          
          $g_{k}, (k=1,...,4)$ are some functions, in dependence on situation, they can be complex or 
          real and we wish they were {\em arbitrary} functions, but it can change in some cases,  
          $a_{\mu}x^{\mu}=a_{1}x^{1}+...+a_{4}x^{4}$, $x^{\mu}$ are the independent variables. We assume that 
          $g_{k} \in \mathcal{C}^{2}, (k=1,...,4)$, (of course, we assume $g_{k}$ are differentiable functions). In the cases of second heavenly equation and mixed 
          heavenly equation we will extend the ansatz (\ref{rozw_suma}) to the functional series.\\         
         Of course, all equations, homogeneous with respect to the derivatives, can be solved by using
         decomposition method (if there exists at least one solution of the determining algebraic system). However, it is possible that found solutions of determining algebraic system will determine the classes of the 
         solutions, which are useless from the physical viewpoint. So, the problem of finding of solutions
         of given equation is reduced to the problem of solving of the determining algebraic system.
         
         It should be also mentioned here that the first method of finding of functionally invariant solutions, applied to the wave equation, comes from \cite{sobolew}, but without the idea of decomposition method, introduced for the first time in \cite{stepien2010} and applied in this paper. In \cite{erugin} and \cite{erugin_smirnow} some extension (obtained by using a method, called also, as Erugin's method, \cite{Meleshko}), of the results obtained in \cite{sobolew}, has been presented. In \cite{menszik1972_I},
         the extension of this above mentioned method, was presented and applied for nonlinear partial differential equations of second order and in \cite{menszik1975} some analogical results were obtained for some kind of quasilinear partial differential equations of second order.
         
         However, the method of searching of the solutions of the form (\ref{ans1}), introduced, for the first time, in \cite{stepien2010}, "looks" at the investigated nonlinear partial differential equation, by the wievpoint of homogenity of some fragments of this equation, with respect to the derivatives and so it differs from these methods mentioned above. By comparison with the methods mentioned above, it seems to be more simple than they. Moreover, we have stated above that decomposition method can be applied for partial differential equation of arbitrary order, if this equation can be decomposed on proper fragments, mentioned above.
          In \cite{stepien2009} some extension of this method (we apply its version in the current paper) was 
          presented.

  \section{Heavenly equations}

   \subsection{Complex Monge-Amp$\grave{e}$re equations}
   
   The Einstein vacuum equation in the complex four-dimensional Riemann space together with the constraint of (anti-)self-duality can be reduced to the complex Monge-Amp$\grave{e}$re equation, \cite{plebanski1975}:

   \begin{equation}
   \Omega_{,pr} \Omega_{,qs} - \Omega_{,qr} \Omega_{,ps} = 1. \label{monge_ampere}
   \end{equation}
   
   The metric, corresponding to this equation, is the following, \cite{plebanski1975}:  
   
   \begin{equation}
   ds^{2}=\Omega_{,pr} dp dr + \Omega_{,ps} dp ds + \Omega_{,qr} dq dr + \Omega_{,qs} dq ds,  
   \label{metryka_monge}
   \end{equation} 
   
   where $p,q,r,s \in \mathbb{C}$ and $\Omega(p,q,r,s) \in \mathbb{C}$. 
   
   Because of physical requirements, we limit our considerations to the case: $\Omega(p,q,r,s)=v$, $v \in 
   \mathbb{R}$, ($p,q,r,s \in \mathbb{C}$).  If we choose: $p=z^{1}, q = z^{2}, r=\theta 
   \bar{z}^{1}, s=\bar{z}^{2}$, then, the equation (\ref{monge_ampere}) becomes, \cite{malykh2004}:
   
   \begin{equation}
   v_{,z^{1} \bar{z}^{1}} v_{,z^{2} \bar{z}^{2}} - v_{,z^{1} \bar{z}^{2}} v_{,z^{2} \bar{z}^{1}} = \theta, 
   \label{monge_ampere_rzecz}
   \end{equation}
  
  where $\theta=\pm 1$, $\bar{z}^{1}$ is complex conjugation of $z^{1}$, $v_{,z^{1}}=\frac{\partial 
  v}{\partial z^{1}}$ etc. The metric (\ref{metryka_monge}) has the form, \cite{malykh2004}: 
   
   \begin{equation}
   ds^{2}= v_{,z^{1} \bar{z}^{1}} dz^{1} d\bar{z}^{1} + v_{,z^{1} \bar{z}^{2}} dz^{1} d\bar{z}^{2} + v_{,z^{2} 
   \bar{z}^{1}} dz^{2} d\bar{z}^{1} + v_{,z^{2} \bar{z}^{2}} dz^{2} d\bar{z}^{2}. \label{metryka_rzecz}
   \end{equation} 
  
   If $\theta= 1$, then the equation (\ref{monge_ampere_rzecz}) is called as the elliptic complex Monge-Amp$\grave{e}$re equation and
   if $\theta= -1$, then the equation (\ref{monge_ampere_rzecz}) is called as the hyperbolic complex Monge-Amp$\grave{e}$re equation.
   
  \subsubsection{Elliptic complex Monge-Amp$\grave{e}$re equation}
  
   As we stated it above, elliptic complex Monge-Amp$\grave{e}$re equation has the form:
    
    \begin{equation}
    v_{z^{1} \bar{z}^{1}} v_{z^{2} \bar{z}^{2}} - v_{z^{1} \bar{z}^{2}} v_{\bar{z}^{1} z^{2}} = 1  \label{cma}
    \end{equation} 
  
    This equation has many applications in mathematics and physics, among others, as we stated it in the previous section, 
    equation (\ref{cma}) is strictly connected to instanton solutions of the Einstein equations of gravitational field.
    These solutions are desrcibed by 4-dimensional K\"{a}hler metrics, \cite{malykh2003}: 
    
    \begin{equation}
    ds^{2} = v_{z^{i} \bar{z}^{k}} dz^{i} d\bar{z}^{k}, \label{metr_kahler}
    \end{equation} 
    
    where we sum over the two values of both: unbarred and barred indices and $v_{z^{i} \bar{z}^{k}}=\frac{\partial^{2} v}{\partial z^{i} 
    \partial \bar{z}^{k}}$. 
    
    The metric satisfies the vacuum Einstein equations of gravitational field with Euclidean signature, provided that the K\"{a}hler 
    potential is some solution of (\ref{cma}). We will look for non-invariant, real solutions of (\ref{cma}), which can be used for 
    construction of hyper K\"{a}hler metrics, not possesing any Killing vectors. One of them is the $K$3 surface (Kummer surface), being
    the most important gravitational instanton, \cite{malykh2003}, \cite{atyiah_hitchin_singer}. 
    In \cite{malykh2003} some exact, non-invariant and real solution of (\ref{cma}) was found, by some reduction of the problem of solving 
    (\ref{cma}) to solving some linear system of equations. Namely, this solution has the form:
    
    \begin{equation}
    \begin{gathered}
    w = \sum^{\infty}_{k=-\infty} \exp{\{2 \Im([A^{2}_{k}(B^{2}_{k}+1)+1]z^{2})\}} \{ \exp[2B_{k} \Re{[A_{k}(p+\gamma z^{2})]}]\\
    \times \Re{\{ C_{k} \exp{[2i [\Im{(A_{k}(p+\gamma z^{2}))} - 2B_{k} \Re{(A^{2}_{k} z^{2})}]]}\}}\\
    + \exp{[-2B_{k} \Re{[A_{k}(p+\gamma z^{2})]}]} \Re{\{ H_{k} \exp{[2i[\Im{[A_{k}(p+\gamma z^{2})]}+2B_{k} \Re{(A^{2}_{k} z^{2})}]]}\}\}},
    \end{gathered}
    \end{equation} 
    
    where $A_{k}, C_{k}, H_{k}$ are arbitrary complex constants, $B_{k} = \sqrt{1-1/\mid A_{k} \mid^{2}}$, $\gamma$ is arbitrary
    real constant, $w = e^{-\psi}$, and $\psi$ is a solution of Legendre transform of elliptic complex Monge-Amp$\grave{e}$re equation, \cite{malykh2003}:
    
    \begin{gather}
    \psi_{p\bar{p}}\psi_{z^{2}\bar{z}^{2}} - \psi_{p\bar{z}^{2}} \psi_{\bar{p}z^{2}}= \psi_{pp} \psi_{\bar{p}\bar{p}} - \psi^{2}_{p\bar{p}},
    \end{gather}
   
   and:
   
    \begin{gather}
    v = \psi - p \psi_{p} - \bar{p} \psi_{,\bar{p}}, \hspace{0.08 in} v_{z^{1}} = p, \hspace{0.08 in} v_{\bar{z}^{1}} = \bar{p}. 
    \label{leg_cma}
    \end{gather} 
    
    Some other solutions, functionally invariant ones, have been found in \cite{malykh2011}:
    
    \begin{gather}
    w = \int^{a_{1}}_{a_{0}} F(a, \beta_{a} + i \delta_{a}) da + \sum_{k} F_{k} (\beta_{a_{k}} + i 
    \delta_{a_{k}}) + c.c.,
    \end{gather}
    
    where $a \in \mathbb{R}$ and:
    
    \begin{equation}
    \begin{gathered}
    v = w - p w_{,p} - \bar{p} w_{,\bar{p}} - r w_{,r}, \hspace{0.05 in} z^{1} = -w_{,p}, 
    \hspace{0.05 in} \bar{z}^{1} = - w_{,\bar{p}}, \\
    \rho = \xi + \bar{\xi} = -w_{,r}, \hspace{0.05 in}
    \beta_{a_{k}} = p + \bar{p} + i a(\bar{p} - p), \label{malykhSIGMA}\\ 
    \delta_{a_{k}} = i \sqrt{\gamma}\bigg(r + \frac{a+i}{a-i} z 
    + \frac{a- i}{a+i} \bar{z}\bigg), \hspace{0.05 in}  \gamma = a^{2} + 1 
    \end{gathered}
    \end{equation}
    
    and $\xi, \bar{\xi}$ are parameters of the symmetry group of (\ref{cma}).

   \subsubsection{Hyperbolic complex Monge-Amp$\grave{e}$re equation transformed by Legendre transformation}

   After applying Legendre transformation, \cite{malykh2004}:
   
  \begin{equation}
 w = v - z^{1} v_{,z^{1}} - \bar{z}^{1} v_{,\bar{z}^{1}}, \hspace{0.05 in}  p = v_{,z^{1}}, \hspace{0.05 in}  \bar{p} = v_{,\bar{z}^{1}}, \hspace{0.05 in} z^{1}=-w_{,p}, \hspace{0.05 in} \bar{z}^{1}=-w_{,\bar{p}} \hspace{0.05 in} , \label{legendre_monge_ampere}
     \end{equation}

    the hyperbolic complex Monge-Amp$\grave{e}$re equation (\ref{monge_ampere_rzecz}) becomes (if $\theta=-1$), \cite{malykh2004}:
    
    \begin{equation}
 w_{,p\bar{p}} w_{,z^{2}\bar{z}^{2}} - w_{,p\bar{z}^{2}} w_{,\bar{p} z^{2}} - w^{2}_{,p\bar{p}} + w_{,pp} w_{,\bar{p}\bar{p}} = 0 . \label{monge_ampere_legendre}
    \end{equation}
   
   The metric (\ref{metryka_rzecz}), governed by (\ref{monge_ampere_rzecz}), after applying the transformation  (\ref{legendre_monge_ampere}), has the form, \cite{malykh2004}:
   
   \begin{equation}
   \begin{gathered}
   ds^{2} = \frac{1}{(w_{,pp}w_{,\bar{p}\bar{p}} - w^{2}_{,p\bar{p}})} \bigg[w_{,pp}(w_{,p\bar{p}} dp + w_{,\bar{p}z^{2}}dz^{2})^{2} + w_{,\bar{p}\bar{p}} (w_{,p\bar{p}} d\bar{p} + w_{,p\bar{z}^{2}} d\bar{z}^{2})^{2} + \\ 
   \frac{w_{,pp}w_{,\bar{p}\bar{p}} + w^{2}_{,p\bar{p}}}{w_{,p\bar{p}}} \mid w_{,p\bar{p}} dp + w_{,\bar{p} z^{2}} dz^{2} \mid^{2} \bigg] - \frac{w_{,pp}w_{,\bar{p}\bar{p}} - w^{2}_{,p\bar{p}}}{w_{,p\bar{p}}} dz^{2} d\bar{z}^{2} .  \label{metr_monge_leg} 
   \end{gathered}
   \end{equation}
  
 The condition of existence of Legendre transformation (\ref{legendre_monge_ampere}) has the form: 
  
  \begin{equation}
  w_{,pp}w_{,\bar{p}\bar{p}} - w^{2}_{,p\bar{p}} \neq 0 \label{warunek_monge_ampere}
  \end{equation}
  
  and it must be satisfied for the given solution or class of solutions of (\ref{monge_ampere_legendre}).

  \subsubsection{Non-invariance of the solutions of the hyperbolic complex Monge-Amp$\grave{e}$re equation}\label{niezm_monge} 
   
    As it was showed in \cite{malykh2004}, the conditions of non-invariance of the solutions of the hyperbolic complex Monge-Amp$\grave{e}$re equation are strictly determined by Killing equations.

   The condition equivalent to the Killing equation, has the form (after applying the invertible point transformation, generated by Legendre transformation (\ref{legendre_monge_ampere})), \cite{malykh2004}: 
   
   \begin{equation}
   \begin{gathered}
    p \xi^{1}(-w_{,p},z^{2})+w_{,z^{2}}\xi^{2}(-w_{,p},z^{2}) + 
    \bar{p}\xi^{\bar{1}}(-w_{,\bar{p}},\bar{z}^{2})+w_{,\bar{z}^{2}}\xi^{\bar{2}}(-w_{,\bar{p}},\bar{z}^{2})=\\
    h(-w_{,p},z^{2})+\bar{h}(-w_{,\bar{p}},\bar{z}^{2}).   \label{punktowo_Killing}
   \end{gathered} 
   \end{equation}
   
   The Killing vector exists for given solution of the hyperbolic complex Monge-Amp$\grave{e}$re equation, 
   only if this solution satisfies (\ref{punktowo_Killing}) and then, such solution is invariant. 
   
   In \cite{malykh2004}, the searching of solutions of hyperbolic Monge-Amp$\grave{e}$re equation, has been 
   reduced to solving some two systems of linear partial differential equations, by applying 
  method of partner symmetries.  
  The solutions of these mentioned systems of linear equation, found in \cite{malykh2004}, have the following form:
   
   \begin{equation}
   w = \sum^{n}_{j=1} \alpha_{j} e^{\Sigma_{j}}, \label{rozw_eksp_monge_ampere}
   \end{equation}
   
   where $\Sigma_{j} = \gamma_{j}p + \bar{\gamma}_{j}\bar{p} + \delta_{j} z^{2} + 
   \bar{\delta}_{j} \bar{z}^{2}$. The coefficients $\alpha_{j} \in \mathbb{R}$ are arbitrary, but $\gamma_{j}, 
   \delta_{j}$ must satisfy the following relations, \cite{malykh2004}: 
   
   \begin{enumerate}
   
   \item for the first system : \\
   
   \begin{equation}
   \mid \gamma_{j} \mid^{2} = a \gamma_{j} + \bar{a} \bar{\gamma_{j}}, \hspace{0.4 in}  \delta_{j} = i 
   \frac{\gamma^{2}_{j} - (\bar{a}+i\bar{b}) \gamma_{j}}{\bar{a}}, \label{wspI}
   \end{equation}

   \item for the second system:\\
   
   \begin{equation}
   \begin{gathered}
   \delta_{j}=\bigg(\nu + i - i \frac{\gamma_{j}}{{\bar{\gamma}_{j}}}\bigg) \gamma_{j}, \\
   c.c., \label{wspII}
   \end{gathered}
   \end{equation} 
   
   where $\nu=const$, and $a, b$ - arbitrary complex constants.
   
   \end{enumerate} 
   
   \vspace{0.2 in}
   
  So, there are two solutions of the hyperbolic complex Monge-Amp$\grave{e}$re equation and they are non-invariant, if $n \geq 4$, because they do not satisfy Killing equation (\ref{punktowo_Killing}), \cite{malykh2004}. Namely, it is provided by the fact, that the matrix of coefficients, \cite{malykh2004}: 
  
   \begin{equation}
          M = \left( \begin{array}{cccc}
          1 & e^{-2i\varphi_{1}} & e^{2i\varphi_{1}} & e^{-4i\varphi_{1}} \\
          1 & e^{-2i\varphi_{2}} & e^{2i\varphi_{2}} & e^{-4i\varphi_{2}} \\
          1 & e^{-2i\varphi_{3}} & e^{2i\varphi_{3}} & e^{-4i\varphi_{3}} \\
          1 & e^{-2i\varphi_{4}} & e^{2i\varphi_{4}} & e^{-4i\varphi_{4}}
      \end{array} \right ), \label{macierz_monge_ampere}
          \end{equation} 
   
   \vspace{0.2 in} 
   
   where $\varphi_{j} =\arg{(\gamma_{j})}$, $(j=1,...,4)$, is non-singular. Hence:
         
         \begin{equation}
            \Sigma_{j}= \gamma_{j}p + \bar{\gamma}_{j}\bar{p} + \delta_{j} z^{2} + \bar{\delta}_{j} \bar{z}^{2}
        \end{equation}
   
   are linearly independent and the transformations from $p, \bar{p}, z^{2}, \bar{z}^{2}$ to $\Sigma_{j}$ are invertible,  \cite{malykh2004}. So, as it has been proved in  \cite{malykh2004}, after inserting each of these above solutions into Killing equation (\ref{punktowo_Killing}), this equation becomes into:
   
   \begin{equation}
     F_{1}(\Sigma_{1},\Sigma_{2}, \Sigma_{3}, \Sigma_{4})=0. \label{killing_po}
   \end{equation}

   Hence, as it has been proved in \cite{malykh2004}, the equation (\ref{punktowo_Killing}) cannot be satisfied identically for the solution of Legendre-transformed hyperbolic complex Monge-Amp$\grave{e}$re equation (\ref{monge_ampere_legendre}), found in \cite{malykh2004}.

   As it has been pointed out in \cite{malykh2004}, the solutions, given by (\ref{rozw_eksp_monge_ampere}) and (\ref{wspI}) can be 
   generalized thanks to the functional invariance, established in a theorem proved in \cite{malykh2004}. 
   According to this theorem, we have that:
   
   \begin{equation}
   w = f\bigg(\sum^{n}_{j=1} \alpha_{j} e^{\Sigma_{j}} \bigg ), \label{rozw_f_eksp_monge_ampere}
   \end{equation}
     
     where the coefficients in $\Sigma_{j}$ satisfy (\ref{wspI}) and $f$ is arbitrary function, 
     ($f \in \mathcal{C}^{2}$), is also the class of the solutions of one of the mentioned above systems of 
     linear PDE's and of course, of hyperbolic complex Monge-Amp$\grave{e}$re equation 
     (\ref{monge_ampere_legendre}). So, (\ref{rozw_f_eksp_monge_ampere}) is some functionally invariant solution 
     of hyperbolic complex Monge-Amp$\grave{e}$re equation (\ref{monge_ampere_legendre}). 
   
   \subsection{Second heavenly equation of Pleba\'{n}ski}\label{niezm_second_heav}

   The second heavenly equation of Pleba\'{n}ski has the following form, \cite{plebanski1975}, 
   \cite{abraham2007}, \cite{malykh2004}:
   
   \begin{equation}
     v_{,xx} v_{,yy} - v^{2}_{,xy} + v_{,xw} + v_{,yz} = 0,   \label{heavenly} 
    \end{equation} 
       
   where $v(x,y,w,z) \in \mathbb{C}$ is a holomorphic function, $v_{xx}=\frac{\partial^{2} v}{\partial x^{2}}$ 
   etc. and $x, y, w, z \in \mathbb{C}$. 
   The heavenly metric has the form, \cite{abraham2007}, \cite{malykh2004}:
   
   \begin{equation}
     ds^{2} = dw dx + dz dy - v_{,xx} dz^{2}  - v_{,yy} dw^{2} + 2 v_{,xy} dw dz. 
     \label{metryka}
   \end{equation}

      In \cite{malykh2004} the second heavenly equation (\ref{heavenly}) has been 
      transformed by partial Legendre transformation:

      \begin{gather}
         \vartheta = v  - w v_{,w} - y v_{,y}, \hspace{0.2 in} v_{,w} = t, \\ 
         v_{,y} = r, \hspace{0.2 in} w = -\vartheta_{,t}, \hspace{0.2 in} y =-\vartheta_{,r}.  \label{leg} 
       \end{gather}

     We need also to remember that this above transformation exists, if the following condition is satisfied, 
     \cite{malykh2004}:
     
      \begin{equation}
          \vartheta_{,tt} \vartheta_{,rr} - (\vartheta_{,rt})^{2} \neq 0.  \label{warunek_h} 
       \end{equation}

    Second heavenly equation, transformed by Legendre transformation (\ref{leg}), has the form, \cite{malykh2004}:
     
     \begin{equation}
      \vartheta_{,tt} (\vartheta_{,xx} + \vartheta_{,rz}) + \vartheta_{,xt}(\vartheta_{,rr} - 
      \vartheta_{,xt}) - \vartheta_{,rt}(\vartheta_{,rx} + \vartheta_{,tz}) = 0 .    \label{heavenly_leg}
     \end{equation} 
     
     The metric (\ref{metryka}) transformed by Legendre transformation, has the form, \cite{malykh2004}:

     \begin{equation}
     \begin{gathered}
      ds^{2} = \frac{1}{\vartheta_{,tt}(\vartheta_{,tt} \vartheta_{,rr} - \vartheta^{2}_{,tr})}\\ 
			(\vartheta_{,tt}(\vartheta_{,tt} dt + \vartheta_{,tr} dr + \vartheta_{,tx} dx +  
      \vartheta_{,tz} dz) + \\
			(\vartheta_{,tt} \vartheta_{,rx} - \vartheta_{,tr} 
      \vartheta_{,tx})dz)^{2} -
       \frac{\vartheta_{,tt} \vartheta_{,xx} - \vartheta^{2}_{,tx}}{\vartheta_{,tt}} dz^{2} -  \\
       (\vartheta_{,tt}dt + \vartheta_{,tr} dr + \vartheta_{,tx} dx + \vartheta_{,tz} dz) dx  \\
       - (\vartheta_{,rt} dt + \vartheta_{,rr} dr + \vartheta_{,rx} dx + \vartheta_{,rz} dz) dz ,  
        \label{metryka_leg} 
      \end{gathered}
      \end{equation} 
      
      where $\vartheta(x,r,t,z)$ is the potential, which satisfies Legendre transformed second heavenly 
      equation (\ref{heavenly_leg}).

     In the aim of linearization of the above equation, there in \cite{malykh2004}, translational symmetries 
     have been applied.
	 In the case of so called equal symmetries, instead of the equation (\ref{heavenly_leg}), the following system of equations has been investigated, \cite{malykh2004}:

	 \begin{gather}
	   \vartheta_{,rt} + \vartheta_{,rr} - \vartheta_{,xt} = 0,  \label{eq_symm1} \\
	   \vartheta_{,xx} + \vartheta_{,rz} = 0, \label{eq_symm2} \\
	   \vartheta_{,rx} + \vartheta_{,xt} + \vartheta_{,tz} = 0. \label{eq_symm3}
	  \end{gather}

	  In the case of so called higher symmetry, the following system has been considered, instead of 
	  (\ref{heavenly_leg}), \cite{malykh2004}:
	  
	   \begin{gather}
	     \vartheta_{,rr} - \vartheta_{,xt} = 0, \label{high_symm1}  \\ 
		 \vartheta_{,rx} + \vartheta_{,tz} = 0, \label{high_symm2} \\
		 \vartheta_{,xx} + \vartheta_{,rz} = 0.  \label{high_symm3} 
           \end{gather}
	  
	  The appropriate form of the Killing equation for second-havenly equation, transformed by Legendre transformation was derived in  
     \cite{malykh2004}.

       In \cite{malykh2004}, a solution of the equations (\ref{eq_symm1})-(\ref{eq_symm3}) 
       and (\ref{high_symm1})-(\ref{high_symm3}) have been obtained:

       \begin{equation}
       \vartheta = \sum^{n}_{j=1} m_{j} \exp{(\alpha_{j} t + \gamma_{j} r + \zeta_{j} x + \lambda_{j} z)},  
       \label{rozw_malykh2004}  
       \end{equation}

       where the coefficients must satisfy the relations, \cite{malykh2004}:
       
         \begin{enumerate}
         \item for the system (\ref{eq_symm1})-(\ref{eq_symm3}):
         
             \begin{equation}
               \alpha_{j} = \frac{\gamma^{2}_{j}}{\zeta_{j} - \gamma_{j}}, \hspace{0.05 in} 
               \lambda_{j}=-\frac{\zeta^{2}_{j}}{\gamma_{j}}, \label{w_rownsym}
             \end{equation}
        
        \item for the system (\ref{high_symm1})-(\ref{high_symm3}):
        
            \begin{equation} 
               \alpha_{j} = \frac{\gamma^{2}_{j}}{\zeta_{j}}, \hspace{0.05 in} 
               \lambda_{j}=-\frac{\zeta^{2}_{j}}{\gamma_{j}} . \label{w_wyssym}
            \end{equation} 
            \end{enumerate} 
       
		Hence, these above solutions are non-invariant, if $n \geq 4$, then, they generate metrics without Killing vector. It is provided by the fact that as in the case of the solutions of hyperbolic complex Monge-Ampere equation, the matrices of coefficients for solutions given either by (\ref{rozw_malykh2004}) and (\ref{w_rownsym}) or by (\ref{rozw_malykh2004}) and (\ref{w_wyssym}), are non-singular, \cite{malykh2004}.

	 \subsection{Mixed heavenly equation}
 
 There in \cite{sheftel2009}, has been derived and investigated, so called mixed heavenly equation, which, after symmetry reduction, has the 
 form:
 
 \begin{equation}
 v_{,ty} v_{,xz} - v_{,tz} v_{,xy} + v_{,tt} v_{,xx} - v^{2}_{,tx} = \theta , \label{mixed}
 \end{equation}
 
 where $\theta = \pm 1$.

 After making Legendre transformation, \cite{sheftel2009}:
 
   \begin{equation}
   \begin{gathered}
  p=v_{,x}, \hspace{0.1 in} q =v_{,z}, \hspace{0.1 in} w(p,q,t,y)=v-xv_{,x}-zv_{,z},\\
  x=-w_{,p}, \hspace{0.1 in} z = -w_{,q} \hspace{0.05 in}, \label{legendre_tr}
  \end{gathered}
  \end{equation} 
  
  the equation (\ref{mixed}) becomes, \cite{sheftel2009}: 
  
  \begin{eqnarray}
	\begin{gathered}
  w_{,tq} w_{,py} - w_{,pq} w_{,ty} + w_{,tt} w_{,qq} - w^{2}_{,tq} + \\ 
	\theta(w_{,pp} w_{,qq} - (w_{,pq})^{2}) = 0. \label{leg_mixed}
	   \end{gathered}
  \end{eqnarray}

  The condition of existence of Legendre transformation (\ref{legendre_tr}) has the form, \cite{sheftel2009}:
  
  \begin{equation}
    w_{,pp} w_{,qq} - (w_{,pq})^{2} \neq 0 .   \label{warunek_mix} 
  \end{equation}

 The problem of obtaining non-invariant solutions of (\ref{leg_mixed}), has been reduced in \cite{sheftel2009}, to investigation of the following set of linear equations (for $\theta = 1$): 
  \begin{eqnarray}
  w_{,\eta \eta} + w_{,\xi \xi} - w_{,\xi q} = 0, \label{heav_lin1} \\
  w_{,\xi q} - w_{,\eta q} + w_{,\xi y} = 0, \\
  w_{,\xi q} + w_{,\eta q} - w_{,qq} + w_{,\eta y} = 0, \label{heav_lin3}
   \end{eqnarray}
   
   where $\eta=p+t, \xi=p-t$. 
   
   One of the solutions of the system (\ref{heav_lin1})-(\ref{heav_lin3}), are, \cite{sheftel2009}:
   
     \begin{equation}
     \begin{gathered}
    w=\sum_{j} \exp{\bigg(\pm \sqrt{A_{j}(A_{j}-B_{j})}(\eta+\frac{B_{j}}{A_{j}}y)\bigg)}\\ 
    \times \{C_{j} \cos{(A_{j}\xi+B_{j}(q-y))} + H_{j} \sin{(A_{j}\xi+B_{j}(q-y))}\}.  \label{rozw_sheftel1}
    \end{gathered}
    \end{equation}
   
   where: $A_{j}, B_{j}, C_{j}, H_{j}$ are arbitrary constants and $\eta=p+t, \xi=p-t$.
     
     This above solution is non-invariant solution, because it depends on four independent combinations of the variables $\eta, \xi, q, y$,
     \cite{sheftel2009}.

	 \subsection{Asymmetric heavenly equation}  
	  
	  In \cite{sheftel2009}, so called asymmetric heavenly equation has been derived:
	  
	  \begin{equation}
	    u_{,tx} u_{,ty} - u_{,tt} u_{,xy} + Au_{,tz} + Bu_{,xz} + Cu_{,xx} = 0. \label{asymm}
	   \end{equation}
	  
	 When $B=0$, then, this above equation is called as evolution form of the second heavenly equation, 
	 \cite{sheftel2009}, \cite{Finley_Plebanski_etal}, \cite{Plebanski_Przanowski}, \cite{Boyer_Plebanski}.  
	 
	   \subsection{General heavenly equation}
    This equation was derived in \cite{Doubrov_Ferap} and it was investigated in \cite{malykh2011} and in \cite{malykh2011JPA}. It has  
		     the form, \cite{malykh2011JPA}:

    \begin{eqnarray}
		\begin{gathered}
    \alpha \omega_{,z^{1}z^{2}} \omega_{,z^{3}z^{4}} + \beta \omega_{,z^{1}z^{3}} \omega_{,z^{2}z^{4}} + \gamma \omega_{,z^{1}z^{4}} 
		\omega_{,z^{2}z^{3}} = 0,  \\ 
		\omega = \omega(z^{1}, z^{2}, z^{3}, z^{4})  \label{genHeav}
    \end{gathered}
		\end{eqnarray}

    where

    \begin{eqnarray}
    \alpha + \beta + \gamma = 0,\\
    m_{1} = \omega_{,z^{1}z^{3}} \omega_{,z^{2}z^{4}} - \omega_{,z^{1}z^{4}} \omega_{,z^{2}z^{3}} \neq 0 
    \label{warunek_m1}
    \end{eqnarray}

    There the following theorem was presented,  \cite{malykh2011JPA}, namely that if $Q(z^{1}, z^{2}, z^{3}, z^{4})$ satisfies 
    (\ref{genHeav}) and the differential constraint:

     \begin{equation}
     \begin{gathered}
    \alpha (Q_{,z^{1}} Q_{,z^{2}} Q_{,z^{3} z^{4}}  + Q_{,z^{3}} Q_{,z^{4}} Q_{,z^{1} z^{2}}) + \beta  (Q_{,z^{1}}   
    Q_{,z^{3}} Q_{,z^{2} z^{4}}  + \\ 
		Q_{,z^{2}} Q_{,z^{4}} Q_{,z^{1} z^{3}}) + \label{constraint} 
    \gamma (Q_{,z^{1}} Q_{,z^{4}} Q_{,z^{2} z^{3}}  + Q_{,z^{2}} Q_{,z^{3}} Q_{,z^{1} z^{4}}) = 0,
    \end{gathered}
    \end{equation}

    then the function $\omega(z^{1}, z^{2}, z^{3}, z^{4})$, determined {\em{implicite}} by $R(\omega, Q) = 0$, where $R$ -  
    arbitrary smooth function, is also a solution of (\ref{genHeav}). Hence, it is functionally invariant solution.

    In \cite{malykh2011JPA} real general heavenly equation has been derived:

     \begin{eqnarray}
		 \begin{gathered}
    \alpha \omega_{,z^{1}\bar{z}^{1}} \omega_{,z^{2}\bar{z}^{2}} + \beta \omega_{,z^{1}z^{2}} \omega_{,\bar{z}^{1}\bar{z}^{2}} + \gamma 
		\omega_{,z^{1}\bar{z}^{2}} \omega_{,z^{2}\bar{z}^{1}} = 0,  \\ 
		\omega = \omega(z^{1},   
    \bar{z}^{1},  z^{2}, \bar{z}^{2}),  \label{genHeavReal}
		   \end{gathered}
    \end{eqnarray}

     and 

    \begin{equation}
    m_{2} = \omega_{,z^{1} z^{2}} \omega_{,\bar{z}^{1}\bar{z}^{2}} - \omega_{,z^{1}\bar{z}^{2}} 
    \omega_{,z^{2}\bar{z}^{1}} \neq 0.  \label{warunek_m2}
    \end{equation}

   One can also consider so-called real cross-sections of (\ref{genHeav}), derived in \cite{malykh2011JPA}:

     \begin{eqnarray}
		 \begin{gathered}
    (\delta^{2} + 1) \omega_{,z^{1}\bar{z}^{1}} \omega_{,z^{2}\bar{z}^{2}} - \delta^{2} \omega_{,z^{1}z^{2}} \omega_{,\bar{z}^{1}\bar{z}^{2}} - \omega_{,z^{1}\bar{z}^{2}} \omega_{,z^{2}\bar{z}^{1}} = 0,  \\ 
    \frac{\beta}{\gamma} > 0, \ \ \beta = \gamma \delta^{2}, \ \ \delta > 0 \label{genHeavRealSec1}
		  \end{gathered}
    \end{eqnarray}

    and

    \begin{eqnarray}
	  \begin{gathered}
    (\delta^{2} - 1) \omega_{,z^{1}\bar{z}^{1}} \omega_{,z^{2}\bar{z}^{2}} - \delta^{2} \omega_{,z^{1}z^{2}} \omega_{,\bar{z}^{1}\bar{z}^{2}} + \omega_{,z^{1}\bar{z}^{2}} \omega_{,z^{2}\bar{z}^{1}} = 0,  \\
    \frac{\beta}{\gamma} < 0, \ \ \beta = - \gamma \delta^{2}, \ \ \delta > 0 \label{genHeavRealSec2}
    \end{gathered}
	  \end{eqnarray}

     \section{Some new classes of exact solutions and their non-invariance}

			 In this section we find the classes of exact solutions of the equations,  (\ref{cma}), (\ref{monge_ampere_legendre}),  (\ref{heavenly_leg}), (\ref{leg_mixed}), (\ref{asymm}), (\ref{genHeavReal}), (\ref{genHeavRealSec2}), and we check the non-invariance of theese classes.
     
     \subsection{Class of exact solutions of elliptic complex Monge-Amp$\grave{e}$re equation}
     
     Now we want to find the class of exact solutions of elliptic complex Monge-Amp$\grave{e}$re equation 
     (\ref{cma}). In contrary to the next subsections of this secion, we do not investigate Legendre transform of origin
     equation. Actually, at first sight, one can think that the simplest way of finding wanted class of exact solutions, is applying
     decomposition method to Legendre transform of (\ref{cma}).\\ 
     
     However, in this case, after obtaining the corresponding 
     determining algebraic system, it turns out that finding of the class of exact solutions, which satisfies
     three conditions of: existence of Legendre transformation, non-invariance and reality, simultaneously, is very hard and it seems that it is possible that there are no appropriate solutions of the determining algebraic system. Thus, we apply 
     decomposition method directly to the elliptic complex Monge-Amp$\grave{e}$re equation (\ref{cma}). Obvioulsy, in this case we cannot use 
     directly the ansatz (\ref{rozw_suma}) to this equation, because the main osbtacle, in applying of the original ansatz,
     is the presence of the free term "1" in (\ref{cma}). So, we apply some modification of the ansatz such that after inserting of it into (\ref{cma}),
     some possibility of balancing of the free term will appear (the necessity of balancing of the free term was
     took into consideration in decomposition method in \cite{stepien2009}). 
      We do it by choosing two functions in (\ref{rozw_suma}), as square functions,
     we choose the functions $g_{k}, (k=1, 2)$ to be square functions and the functions $g_{3}$ is arbitrary function of 
     class $\mathcal{C}^{2}$, but $g_{4} = \bar{g}_{3}$, in order to satisfy the condition of reality of the solution. Hence, we apply the ansatz:
     
      \begin{gather}
      v(x^{\mu})=\beta_{1} + (\Sigma_{1})^{2} + (\Sigma_{2})^{2} + g_{3}(\Sigma_{3}) + g_{4}(\Sigma_{4}), \label{rozw_kwadr} 
             \end{gather}
 
             where:

             \begin{equation}
             \begin{gathered} 
             \Sigma_{1} = a_{\mu} x^{\mu} + \beta_{2}, \hspace{0.2 in} \Sigma_{2} = b_{\mu} x^{\mu} + 
             \beta_{3}, \\
                \Sigma_{3} = c_{\mu} x^{\mu} + \beta_{4}, \hspace{0.2 in}
                \Sigma_{4} = d_{\mu} x^{\mu} + \beta_{5}, \label{argumentykwadr} 
             \end{gathered}
             \end{equation}
          
          $\Sigma_{k} \in \mathbb{R}, (k=1,2)$,
          $g_{3} \in \mathbb{C}$ is {\em arbitrary} function of class $\mathcal{C}^{2}$, $g_{4} = \bar{g}_{3}$, 
          $a_{\mu}x^{\mu}=a_{1}x^{1}+...+a_{4}x^{4}$ and $x^{1}=z^{1}, x^{2}=\bar{z}^{1}, x^{3}=z^{2}, x^{4}=\bar{z}^{2}$ 
          are the independent variables.
     Owing to applying decomposition method directly to (\ref{cma}), not to its Legendre transform (\ref{leg_cma}), we avoid the necessity of 
     satisfying of the condition of existence of Legendre transformation.
     After inserting the ansatz (\ref{rozw_kwadr}) into (\ref{cma}) and collecting proper terms, we derive the determining algebraic system, which wanted solutions are:
     
     \begin{gather}
     a_{1}=\bar{a}_{2}, a_{3}=0, a_{4}=0, b_{1}=1, b_{2}=1, b_{3}=\frac{1}{2\bar{a}_{2}}, b_{4}= \frac{1}{2a_{2}},\\
     c_{1}=0, c_{2}=\frac{1}{2\bar{a}_{2}}, c_{3}=0, c_{4}=\bar{d_{3}}, d_{1}=\frac{1}{2a_{2}}, d_{2}=0, d_{4}=0,\\
     \beta_{1}, \beta_{2}, \beta_{3} \in \mathbb{R}, \hspace{0.05 in} \beta_{5}=\bar{\beta}_{4}.  
     \end{gather} 
     
     Hence, the class of solutions of (\ref{cma}) has the form:
     
     \begin{equation}
     \begin{gathered}
     v(z^{1}, \bar{z}^{1}, z^{2}, \bar{z}^{2}) = \beta_{1} + (\bar{a}_{2} z^{1} + a_{2} \bar{z}^{1} + \beta_{2})^{2} +
     \bigg(z^{1} + \bar{z}^{1} + \frac{1}{2\bar{a}_{2}} z^{2} + \frac{1}{2a_{2}} \bar{z}^{2} + \beta_{3} \bigg)^{2} + \label{rozw_cma} \\
     g_{3}\bigg(\frac{1}{2\bar{a}_{2}} \bar{z}^{1} + \bar{d}_{3} \bar{z}^{2} + \beta_{4} \bigg) + 
     \bar{g}_{3}\bigg(\frac{1}{2a_{2}} z^{1} + d_{3} z^{2} + \bar{\beta}_{4} \bigg) .
     \end{gathered}
     \end{equation}

     Now we need to check, whether the solutions, belonging to the found class, are non-invariant.
     We construct the matrix:
     
     \begin{equation}
          M = \left( \begin{array}{cccc}
      \Gamma_{1} \bar{a}_{2}  & \Gamma_{1} a_{2} & 0 & 0 \\
       \Gamma_{2} &  \Gamma_{2} & \frac{\Gamma_{2}}{2 \bar{a}_{2}} & \frac{\Gamma_{2}}{2 a_{2}} \\
         0 & \frac{g'_{3}}{2\bar{a}_{2}} & 0 & g'_{3} \bar{d}_{3} \\
       \frac{\bar{g}'_{3}}{2 a_{2}} &  0 & \bar{g}'_{3} d_{3} & 0
      \end{array} \right),
          \end{equation}
     
     where $\Gamma_{1} =2(\bar{a}_{2} z^{1} + a_{2} \bar{z}^{1} + \beta_{2}), \Gamma_{2}=2z^{1} + 2\bar{z}^{1} + \frac{z^{2}}{\bar{a}_{2}} + 
     \frac{\bar{z}^{2}}{a_{2}} + 2\beta_{3}$ . In order to provide non-invariance of the solutions belonging to the class (\ref{rozw_cma}), the determinant of this above matrix must not vanish:
     
     \begin{equation}
     \begin{gathered}
     \det{(M)} = -\frac{1}{2\bar{a}^{2}_{2} a^{2}_{2}}(\bar{a}_{2} z^{1} + a_{2} \bar{z}^{1} + \beta_{2}) (2z^{1}\bar{a}_{2} a_{2} 
     + 2\bar{z}^{1}\bar{a}_{2} a_{2} + z^{2} a_{2} + \bar{z}^{2} \bar{a}_{2} + 2\beta_{3} \bar{a}_{2} a_{2})
     \label{wyzn_cma} \\
     (4d_{3}\bar{d}_{3} \bar{a}^{2}_{2} a_{2} - d_{3} \bar{a}_{2} - 4\bar{a}_{2} a^{2}_{2} d_{3} \bar{d}_{3} + \bar{d}_{3} a_{2})
     g'_{3} \bar{g}'_{3} \neq  0
     \end{gathered}
     \end{equation} 
      
     Let us assume that $a_{2} \neq 0$. Hence, the solutions belonging to the class (\ref{rozw_cma}),
     depend on four variables and they are non-invariant, in the regions, where simultaneously the conditions: $\bar{a}_{2} z^{1} + a_{2} \bar{z}^{1} + \beta_{2} \neq 0$ and $2z^{1}\bar{a}_{2} a_{2} + 2\bar{z}^{1}\bar{a}_{2} a_{2} + z^{2} a_{2} + \bar{z}^{2} \bar{a}_{2} + 2\beta_{3} \bar{a}_{2} a_{2} \neq 0$ and $g'_{3} \neq 0$, hold.
     
      Of course, this above found class of solutions of the elliptic complex Monge-Amp$\grave{e}$re equation (\ref{cma}), is the class of {\em real} solutions. Obviously, these solutions are not differentiable in $\mathbb{C}$ sense. They are differentiable in $\mathbb{R}$ sense. If we repeat these above computations, but in the variables $x^{1}=\Re{(p)}, x^{2}=\Im{(p)}, x^{3}=\Re{(z^{2})}, x^{4}=\Im{(z^{2})}$, $x^{k} \in \mathbb{R}, (k=1,2,3,4)$, then it turns out that the function, given by (\ref{rozw_cma}), is still the class of exact solutions of elliptic complex Monge-Amp$\grave{e}$re equation (\ref{cma}) and these solutions are still non-invariant.

 Hence, we have proved the following theorem:
 
 \begin{thm}
 The metric (\ref{metr_kahler}) with $v$, being some solution of elliptic complex Monge-Amp$\grave{e}$re equation (\ref{cma}), belonging to the class, given by by (\ref{rozw_cma}), when (\ref{wyzn_cma}) and the conditions: $\bar{a}_{2} z^{1} + a_{2} \bar{z}^{1} + \beta_{2} \neq 0$, $2z^{1}\bar{a}_{2} a_{2} + 2\bar{z}^{1}\bar{a}_{2} a_{2} + z^{2} a_{2} + \bar{z}^{2} \bar{a}_{2} + 2\beta_{3} \bar{a}_{2} a_{2} \neq 0$ and $g'_{3} \neq 0$, are simultaneously satisfied, does not possess Killing vector.
 \end{thm}

     \subsection{Classes of exact solutions of hyperbolic complex Monge-Amp$\grave{e}$re equation}

     In this subsection we look for the class of exact solutions of hyperbolic complex Monge-Amp$\grave{e}$re equation 
     (\ref{monge_ampere_legendre}). To this effect, we apply directly the ansatz (\ref{rozw_suma}) for 
     (\ref{monge_ampere_legendre}), in contrast with \cite{malykh2004}, where this equation was linearized
     and the obtained systems of linear equations were solved.
     Here: $u \equiv w$ and the independent variables are: $x^{1} = p, x^{2} = \bar{p}, x^{3} = z^{2}, x^{4} = \bar{z}^{2}$, and $g_{j}$ ($j = 1, 2, 3, 4$), are such functions of their arguments that $w(p,\bar{p},z^{2},\bar{z}^{2}) \in \mathbb{R}$. 
  We search the class of non-invariant solutions of (\ref{monge_ampere_legendre}), given by (\ref{rozw_suma}) and (\ref{argumenty1}),  which satisfy (\ref{warunek_monge_ampere}).\\ 
  After making the procedure described in the section 2, we obtain some system of nonlinear algebraic equations, so called determining  algebraic system. Apart from satisfying of it, we require also satisfying of the following conditions: the condition (\ref{warunek_monge_ampere}), the condition of non-singularity of Jacobian matrix and the condition that the solution must be {\em real}. 
  
 We found three classes of non-invariant, exact solutions of (\ref{monge_ampere_legendre}), satisfying the mentioned conditions. These classes are given by (\ref{rozw_suma}) and by the following sets of  relations between the coefficients:\\

 \begin{enumerate}
 \item for the class I
 \begin{equation}
 \begin{gathered}
 a_{1}=a_{2}, a_{3}=-\frac{a_{2}(d_{1}d_{2}-d^{2}_{2}-d_{1}d_{4})}{d_{1}d_{2}}, 
 a_{4}=-\frac{a_{2}(d_{1}d_{2}-d^{2}_{2}-d_{1}d_{4})}{d_{1}d_{2}}, \\
 b_{1}=d_{2}, b_{2}=d_{1} , b_{3}=d_{4}, b_{4} = -\frac{d^{2}_{1}-d^{2}_{2}-d_{1}d_{4}}{d_{2}}, \\
 c_{1}=-iA, c_{2}=iA, c_{3}=-\frac{iA(d_{1}d_{2}+d^{2}_{2}+d_{1}d_{4})}{d_{1}d_{2}}, c_{4}=\frac{iA(d_{1}d_{2}+d^{2}_{2}+d_{1}d_{4})}{d_{1}d_{2}}, \\
 d_{3}=-\frac{d^{2}_{1}-d^{2}_{2}-d_{1}d_{4}}{d_{2}}, \label{rel_monge1} 
 \end{gathered}
 \end{equation} 
  
 in this case: $g_{1}, g_{3} \in \mathbb{R}$, $g_{2} \in \mathbb{C}, g_{4}=\bar{g}_{2}$ (of course $\beta_{5} = \bar{\beta}_{3}$),
  
 \item for the class II
 \begin{equation}
 \begin{gathered}
 a_{1}=0, \hspace{0.05 in} a_{2}=0, \hspace{0.05 in}  a_{3}=\bar{a}_{4},\hspace{0.05 in}  b_{1}=0, 
 \hspace{0.05 in}  b_{2}=0, \hspace{0.05 in} b_{3}=\bar{b}_{4}, \\
 c_{1}=0, \hspace{0.05 in} c_{2}=d_{4}, \hspace{0.05 in} c_{3}=d_{4}, \hspace{0.05 in}
 c_{4}=d_{3}, \hspace{0.05 in} d_{1}=d_{4}, \hspace{0.05 in} d_{2}=0, \label{rel_monge2}
 \end{gathered}
 \end{equation}

   in this case: $g_{1}, g_{2} \in \mathbb{R}$, $g_{3} \in \mathbb{C}, g_{4}=\bar{g}_{3}$ (of course $\beta_{5} = \bar{\beta}_{4}$), 
   
   \item for the class III
   \begin{equation}
   \begin{gathered}
    a_{1}=A_{2}(1+i),  \hspace{0.05 in} a_{2} = \bar{a}_{1},  \hspace{0.05 in} a_{3}=2iA_{2},  
    \hspace{0.05 in} a_{4}=\bar{a}_{3}, \\
    b_{1}=\frac{\sqrt{B^{2}_{3}+B^{2}_{4}}}{2}\bigg(1 + i \frac{B_{3} + B_{4} - 
    \sqrt{B^{2}_{3}+B^{2}_{4}}}{B_{3}-B_{4} + \sqrt{B^{2}_{3}+B^{2}_{4}}}\bigg), \hspace{0.05 in}
    b_{2}=\bar{b}_{1}, \\ 
    b_{3} = B_{3} + i B_{4}, \hspace{0.05 in} b_{4}=\bar{b}_{3}, \hspace{0.05 in} 
    c_{1}=iC_{2}, \hspace{0.05 in} c_{2}=\bar{c}_{1}, \hspace{0.05 in} c_{3}=0, c_{4}=0,\\ 
    d_{1}=H_{2}(-1 + i), \hspace{0.05 in} d_{2}=\bar{d}_{1}, \hspace{0.05 in} d_{3}=2iH_{2}, 
    \hspace{0.05 in} d_{4}=\bar{d}_{3} \label{rel_monge3}
    \end{gathered}
    \end{equation}    
    
    in this case: $g_{j}, (j=1,2,3,4) \in \mathbb{R}$. 
 \end{enumerate}
  
  In all these above cases, $g_{j}, (j=1,2,3,4)$ are the functions of class $\mathcal{C}^{2}$.

  Hence, the found classes of the exact solutions of (\ref{monge_ampere_legendre}) have the form:\\
  
  \begin{enumerate}
  \item class I
  \begin{equation}
  \begin{gathered}
  w(p, \bar{p}, z^{2}, \bar{z}^{2})=\beta_{1} + \\ g_{1}\bigg(a_{2}p+a_{2}\bar{p}-\frac{a_{2}(d_{1}d_{2}-d^{2}_{2}-d_{1}d_{4})}{d_{1}d_{2}}z^{2}-\frac{a_{2}(d_{1}d_{2}-d^{2}_{2}-d_{1}d_{4})}{d_{1}d_{2}}\bar{z}^{2}+\beta_{2}\bigg)+\\
  g_{2}\bigg(d_{2}p+d_{1}\bar{p}+d_{4}z^{2} -\frac{d^{2}_{1}-d^{2}_{2}-d_{1}d_{4}}{d_{2}} \bar{z}^{2}+\beta_{3}\bigg)+\\ g_{3}\bigg(-iAp+iA\bar{p}-\frac{iA(d_{1}d_{2}+d^{2}_{2}+d_{1}d_{4})}{d_{1}d_{2}}z^{2}+\frac{iA(d_{1}d_{2}+d^{2}_{2}+d_{1}d_{4})}{d_{1}d_{2}}\bar{z}^{2}+\beta_{4}\bigg)+\\ 
\bar{g}_{2}\bigg(d_{2}p+d_{1}\bar{p}+d_{4}z^{2} -\frac{d^{2}_{1}-d^{2}_{2}-d_{1}d_{4}}{d_{2}} \bar{z}^{2}+\beta_{3}\bigg), \label{rozw_monge1}
  \end{gathered}
  \end{equation} 

  \vspace{0.2 in}
  
  \item class II 
  \begin{equation}
  \begin{gathered}
 w(p, \bar{p}, z^{2}, \bar{z}^{2}) = \beta_{1} + g_{1}(\bar{a}_{4} z^{2} + a_{4} \bar{z}^{2} + \beta_{2}) + 
 g_{2}(\bar{b}_{4} z^{2} + b_{4} \bar{z}^{2} + \beta_{3}) + \\ 
 g_{3}(d_{4} \bar{p} + d_{4} z^{2} + d_{3} \bar{z}^{2} + \beta_{4}) + \bar{g}_{3}(d_{4} \bar{p} + d_{4} z^{2} + d_{3} \bar{z}^{2} + \beta_{4}), \label{rozw_monge2}
 \end{gathered}
 \end{equation}
 
 \vspace{0.2 in}
 
 \item class III
 \begin{equation}
 \begin{gathered}
  w(p, \bar{p}, z^{2}, \bar{z}^{2}) = \beta_{1} + g_{1}\bigg(A_{2}(1+i)p + A_{2}(1-i) \bar{p} + 2iA_{2} z^{2} -  2iA_{2} \bar{z}^{2} + \beta_{2}\bigg) + \\
 g_{2}\bigg(\frac{\sqrt{B^{2}_{3}+B^{2}_{4}}}{2}\bigg[1 + i \frac{B_{3} + B_{4} - 
    \sqrt{B^{2}_{3}+B^{2}_{4}}}{B_{3}-B_{4} + \sqrt{B^{2}_{3}+B^{2}_{4}}}\bigg]p + \\
    \frac{\sqrt{B^{2}_{3}+B^{2}_{4}}}{2}\bigg[1 - i \frac{B_{3} + B_{4} - 
    \sqrt{B^{2}_{3}+B^{2}_{4}}}{B_{3}-B_{4} + \sqrt{B^{2}_{3}+B^{2}_{4}}}\bigg] \bar{p} + \\
    (B_{3}+iB_{4}) z^{2} + (B_{3}-iB_{4}) \bar{z}^{2} + \beta_{3}\bigg) + \\ 
 g_{3}\bigg(iC_{2}p - iC_{2}\bar{p} + \beta_{4}\bigg) + \\
 g_{4}\bigg(H_{2}(-1+i) p + H_{2}(-1-i) \bar{p} + 2iH_{2} z^{2} - 2iH_{2} \bar{z}^{2} + \beta_{5}\bigg), 
 \label{rozw_monge3} 
 \end{gathered}
 \end{equation}
 
 \end{enumerate} 
  
  where:
  
  \begin{itemize}
  \item for class I: $g_{1}, g_{3}$ are {\em arbitrary} real functions (of class $\mathcal{C}^{2}$) of their arguments, $g_{2}$ is some {\em arbitrary} complex function of its argument and $g_{2} \in \mathcal{C}^{2}$, $\bar{g}_{2}$ is complex conjugation of $g_{2}$, i.e.: $g_{2}=f(\Phi),
  \bar{g}_{2}=f(\bar{\Phi})$, where $\Phi \in \mathbb{C}$ is the argument of $g_{2}$ given in (\ref{rozw_monge1}) and $A, a_{2}, d_{1}, d_{2}, d_{4}, \beta_{1}, \beta_{2}, \beta_{4} \in \mathbb{R}$ and $\beta_{3} \in \mathbb{C}$.
  
  \item for class II:
  $g_{1}, g_{2}$ are {\em arbitrary} real functions (of class $\mathcal{C}^{2}$) of their arguments, $g_{3}$ is some {\em arbitrary} complex function of its argument and $g_{3} \in \mathcal{C}^{2}$, $\bar{g}_{3}$ is complex conjugation of $g_{3}$, i.e.: $g_{3}=f(\Phi),
  \bar{g}_{3}=f(\bar{\Phi})$, where $\Phi \in \mathbb{C}$ is the argument of $g_{3}$ given in (\ref{rozw_monge2}) and $a_{4}, b_{4} \in \mathbb{C}$, $d_{3}, d_{4}, \beta_{k}  \in \mathbb{R}, (k=1,2,3)$ and $\beta_{4} \in \mathbb{C}$.
  
  \item for class III: $g_{j}, (j=1,2,3,4)$ are {\em arbitrary} real functions (of class $\mathcal{C}^{2}$) of their arguments and $A_{2}, B_{3}, B_{4}, C_{2}, H_{2}, \beta_{k} \in \mathbb{R}, (k=1,...,5)$.
  \end{itemize}

 Now, we check, whether the condition (\ref{warunek_monge_ampere}) is satisfied by the solutions belonging to the classes I, II and III. It turns out that it is satisfied, when:

 \begin{enumerate} 
 \item for class I
 \begin{equation}
 \begin{gathered}
 a^{2}_{2}(d_{1}-d_{2})^{2}g''_{1}(\bar{g}''_{2}+g''_{2}) - A^{2}(d_{1}+d_{2})^{2}g''_{3}(\bar{g}''_{2}+g''_{2})+
 (d^{2}_{1}-d^{2}_{2})^{2} g''_{2}\bar{g}''_{2}-\\
 4a^{2}_{2}A^{2}g''_{1}g''_{3} \neq 0, \label{warunek1}
 \end{gathered}
 \end{equation}
 
 \item for class II 
 \begin{equation}
   d^{4}_{4} g''_{3} \bar{g}''_{3} \neq 0, \label{warunek2} 
 \end{equation}

 \item for class III (some algebraic inequality, which we skip in this paper, due to its complicated structure), 
     
 \end{enumerate}

  where $g''_{j}, (j=1,2,3,4)$ denotes second derivative of the function $g_{j}$ with respect to its argument.

 Next, basing on the considerations included in the subsubsection 3.1.2, we make the analysis of non-invariance of the solutions, belonging to the found classes. The Jacobian matrices have the forms:

  \begin{enumerate}
  \item for class I
   
  \begin{equation}
          M = \left( \begin{array}{cccc}
      a_{2} g'_{1} & a_{2} g'_{1} & -a_{2}\frac{d_{1}d_{2}-d^{2}_{2}-d_{1}d_{4}}{d_{1}d_{2}}g'_{1} & 
      -a_{2}\frac{d_{1}d_{2}-d^{2}_{2}-d_{1}d_{4}}{d_{1}d_{2}} g'_{1} \\
       d_{2}g'_{2} &  d_{1}g'_{2} &  d_{4}g'_{2} & - \frac{d^{2}_{1}-d^{2}_{2}-d_{1}d_{4}}{d_{2}} g'_{2} \\
      -i A g'_{3} & i A g'_{3} & - \frac{iA(d_{1}d_{2}+d^{2}_{2}+d_{1}d_{4})}{d_{1}d_{2}}g'_{3} & 
       \frac{iA(d_{1}d_{2}+d^{2}_{2}+d_{1}d_{4})}{d_{1}d_{2}} g'_{3} \\
       d_{1}\bar{g}'_{2} &  d_{2}\bar{g}'_{2} & -\frac{d^{2}_{1}-d^{2}_{2}-d_{1}d_{4}}{d_{2}}\bar{g}'_{2} & 
       d_{4}\bar{g}'_{2}
      \end{array} \right),
          \end{equation}

   \item for class II
          \begin{equation}
          M = \left( \begin{array}{cccc}
           0 & 0 &  \bar{a}_{4}g'_{1} &  a_{4}g'_{1} \\
           0 & 0 &  \bar{b}_{4}g'_{2} &  b_{4}g'_{2}  \\
           0 & d_{4}g'_{3} &  d_{4}g'_{3} &  d_{3}g'_{3} \\
       d_{4}\bar{g}'_{3} & 0 &  d_{3}\bar{g}'_{3} &  d_{4}\bar{g}'_{3}
      \end{array} \right),
          \end{equation}
          
   \item for class III
         \begin{equation}
          M = \left( \begin{array}{cccc}
           A_{2}(1+i)g'_{1} & A_{2}(1-i)g'_{1} & 2iA_{2} g'_{1} & -2iA_{2} g'_{1} \\
           N_{1} g'_{2} & \bar{N}_{1} g'_{2} & (B_{3}+iB_{4}) g'_{2} & (B_{3}-iB_{4}) g'_{2}  \\
           iC_{2} g'_{3} & -iC_{2} g'_{3} & 0 & 0 \\
           H_{2}(-1+i) g'_{4} & H_{2}(-1-i) g'_{4} & 2iH_{2} g'_{4} &  -2iH_{2} g'_{4}
      \end{array} \right),
         \end{equation}  
          
    \end{enumerate}

  where $g'_{j}, (j=1,2,3,4)$ denotes the first derivative of the function $g_{j}$ with respect to its argument
  and $N_{1}=\frac{\sqrt{B^{2}_{3}+B^{2}_{4}}}{2}\bigg(1 + i \frac{B_{3} + B_{4} - 
    \sqrt{B^{2}_{3}+B^{2}_{4}}}{B_{3}-B_{4} + \sqrt{B^{2}_{3}+B^{2}_{4}}}\bigg)$.
  From the requirement of non-vanishing of the determinants of these above matrices, we have:
  
  \begin{enumerate}
  \item for the class I   
  \begin{gather}
    \det{M} = \frac{2ia_{2}A}{d^{2}_{1}d^{2}_{2}}(d^{6}_{1}-d^{6}_{2}-3d^{4}_{1}d^{2}_{2}+3d^{2}_{1}d^{4}_{2})g'_{1}g'_{2}\bar{g}'_{2}g'_{3} \neq 0, \label{wyzn_monge1}
    \end{gather} 
  \item for the class II
  \begin{gather} 
  \det{M} = -(\bar{a}_{4}b_{4}-a_{4}\bar{b}_{4}) d^{2}_{4} g'_{1}g'_{2}g'_{3}\bar{g}'_{3} \neq 0,  
  \label{wyzn_monge2}
  \end{gather}
  \item for the class III
  \begin{gather} 
  \det{M} = 16 A_{2}B_{3}C_{2}H_{2} g'_{1}g'_{2}g'_{3}g'_{4} \neq 0 . \label{wyzn_monge3}
  \end{gather} 
  
  \end{enumerate}

    Let's assume additionally for class I: 
    
    \begin{equation}
    d_{1}d_{2} \neq 0. \label{war2_wyzn_monge}
    \end{equation}

 Let us {\em \bf fix} now the functions $g_{j}, (j=1,...,4)$ in (\ref{rozw_monge1}), (\ref{rozw_monge2}) and (\ref{rozw_monge3}), but such that the conditions (\ref{wyzn_monge1}) (together with (\ref{warunek1}) and (\ref{war2_wyzn_monge})), (\ref{wyzn_monge2}) (together with (\ref{warunek2})) and (\ref{wyzn_monge3}) (together with and $B_{3} - B_{4} + \sqrt{B^{2}_{3} + B^{2}_{4}} \neq 0$ and (A.1)) will be still satisfied, correspondingly. 
 
  We can now repeat from the subsubsection 3.1.2 (basing on \cite{malykh2004}), that the equation (\ref{punktowo_Killing}) cannot be satisfied identically for any solution of Legendre-transformed hyperbolic complex Monge-Amp$\grave{e}$re equation (\ref{monge_ampere_legendre}) just by proper choice of the functions $\xi^{1}, \xi^{\bar{1}}, \xi^{2}, \xi^{\bar{2}}, h, \bar{h}$, because the variables $p, \bar{p}, w_{,z^{2}}, w_{,\bar{z}^{2}}$ explicitly enter into the coefficients of this equation.\\
    $\Sigma_{j}, \hspace{0.05 in} (j=1,2,3,4)$ are linearly independent, for the three above classes of solutions. So, the transformations from $p, \bar{p}, z^{2}, \bar{z}^{2}$ to $\Sigma_{j}$ are invertible and we can express $p, \bar{p}, z^{2}, \bar{z}^{2}$ through $\Sigma_{j}, \hspace{0.05 in} (j=1,2,3,4)$, so that $\Sigma_{j}, \hspace{0.05 in} (j=1,2,3,4)$, can be chosen as new independent variables in (\ref{punktowo_Killing}) and after inserting each of above classes of solutions into the equation (\ref{punktowo_Killing}), this equation  becomes:
  
  \begin{equation}
  F_{3}(\Sigma_{1}, \Sigma_{2}, \Sigma_{3}, \Sigma_{4})=0 .
  \end{equation}

 For example, if we choose $g_{j}=\exp{(\Sigma_{j})}$ in the above found classes of solutions, especially in (\ref{rozw_monge3}), we obtain form similar to the form of the ansatz (\ref{rozw_eksp_monge_ampere}), if $n=4$. 
  So, the solutions, belonging to the above classes: (\ref{rozw_monge1}) or (\ref{rozw_monge2}) or (\ref{rozw_monge3}), with such fixed functions $g_{k}, (k=1,...,4)$, do not have functional arbitrariness. Hence, after taking into account these above arguments, we see that  these solutions  cannot satisfy the first-order Killing equation (\ref{punktowo_Killing}).\\ 
 Thus, we see that the Killing equation (\ref{punktowo_Killing}) cannot be satisfied identically for any solution, belonging to the classes of the form: (\ref{rozw_monge1}) or (\ref{rozw_monge2}) or (\ref{rozw_monge3}).
  
  Of course, all these above found classes of solutions of the hyperbolic complex Monge-Amp$\grave{e}$re equation (\ref{monge_ampere_legendre}), are the classes of {\em real} solutions. Obviously, these solutions are not differentiable in $\mathbb{C}$ sense. They are differentiable in $\mathbb{R}$ sense. If we repeat these above computations, but in the variables $x^{1}=\Re{(p)}, x^{2}=\Im{(p)}, x^{3}=\Re{(z^{2})}, x^{4}=\Im{(z^{2})}$, $x^{k} \in \mathbb{R}, (k=1,2,3,4)$, then it turns out that the functions, given either by (\ref{rozw_monge1}) or by (\ref{rozw_monge1}) or by (\ref{rozw_monge3}), are still the classes of exact solutions of hyperbolic complex Monge-Amp$\grave{e}$re equation (\ref{monge_ampere_legendre}), these solutions are still non-invariant and the condition (\ref{warunek_monge_ampere}) is still satisfied.

  Hence, we have proved the following theorem:

  \begin{thm}
  The metric (\ref{metr_monge_leg}) with $w$, being some solution of hyperbolic complex Monge-Amp$\grave{e}$re equation (\ref{monge_ampere_legendre}), belonging to any class, defined by:
 
  \begin{enumerate}
  \item  
  (\ref{rozw_monge1}) - class I, (when the relations: (\ref{warunek1}),  (\ref{wyzn_monge1}) and 
  (\ref{war2_wyzn_monge}) 
  hold) 
  \item   
  (\ref{rozw_monge2}) - class II, (when the relations: (\ref{warunek2}), (\ref{wyzn_monge2}) hold) 
  \item 
  (\ref{rozw_monge3}) - class III, (when are satisfied the relations: (\ref{wyzn_monge3}), $B_{3} - 
  B_{4} + \sqrt{B^{2}_{3} + B^{2}_{4}} \neq 0$ and some algebraic inequality mentioned above),
  \end{enumerate}
  
  where the functions $g_{j}, (j=1,...,4)$, are fixed, does not possess Killing vector. 
  \end{thm} 
      
	  \subsection{Classes of exact solutions of second heavenly equation} 
	          
        Now, in order to find new classes of non-invariant solutions of heavenly equation, we use decomposition method, in the cases of equal symmetries and higher symmetries. We apply the ansatz (\ref{rozw_suma}) for  the systems (\ref{eq_symm1})-(\ref{eq_symm3}) and (\ref{high_symm1})-(\ref{high_symm3}), but now $u \equiv \vartheta$, the independent variables are: $x^{1} = x, x^{2} = r, x^{3} = t, x^{4} = z$, and $g_{k}$, ($k = 1, 2, 3, 4$), are {\em arbitrary} holomorphic, complex-valued functions of their arguments.    
          The ansatz (\ref{rozw_suma}) presents the class of solutions of the systems 
          (\ref{eq_symm1})-(\ref{eq_symm3}), (\ref{high_symm1})-(\ref{high_symm3}), when      
          the relations, which must be satisfied by the coefficients, are, as follows:

          \begin{enumerate}
          \item for equations (\ref{eq_symm1})-(\ref{eq_symm3}) - the case of equal symmetries:
                 
            \begin{equation}
            \begin{gathered}
              a_{1} = \frac{a_{2}(a_{2}+a_{3})}{a_{3}}, \hspace{0.2 in}
              a_{4} = -\frac{a_{2}(a_{2}+a_{3})^{2}}{a^{2}_{3}}, \\
              b_{1} = \frac{b_{2}(b_{2}+b_{3})}{b_{3}}, \hspace{0.2 in}
              b_{4} = -\frac{b_{2}(b_{2}+b_{3})^{2}}{b^{2}_{3}}, \\
              c_{1} = \frac{c_{2}(c_{2} + c_{3})}{c_{3}}, \hspace{0.2 in}
              c_{4} = - \frac{c_{2}(c_{2} + c_{3})^{2}}{c^{2}_{3}}, \\
              d_{1} = \frac{d_{2}(d_{2} + d_{3})}{d_{3}}, \hspace{0.2 in} 
              d_{4} = - \frac{d_{2}(d_{2} + d_{3})^{2}}{d^{2}_{3}},
              \label{wsp_eq_symm}
            \end{gathered}
            \end{equation}

          \item for equations (\ref{high_symm1})-(\ref{high_symm3}) - the case of higher symmetry:
          
          \begin{itemize}
          \item subclass I
                
                  \begin{equation}
                  \begin{gathered}
                   a_{3} = \frac{a^{2}_{2}}{a_{1}}, \hspace{0.2 in}
                   a_{4} = - \frac{a^{2}_{1}}{a_{2}}, \\
                   b_{3} = \frac{b^{2}_{2}}{b_{1}}, \hspace{0.2 in}
                   b_{4} = -\frac{b^{2}_{1}}{b_{2}}, \\
                   c_{3} = \frac{c^{2}_{2}}{c_{1}}, \hspace{0.2 in}
                   c_{4} = -\frac{c^{2}_{1}}{c_{2}}, \\
                   d_{3} = \frac{d^{2}_{2}}{d_{1}}, \hspace{0.2 in}
                   d_{4} = -\frac{d^{2}_{1}}{d_{2}}, \label{wsp_high_symm1}
                \end{gathered}
                \end{equation}
               
         \item subclass II 
            \begin{equation}
            \begin{gathered}
              a_{1}=\frac{a^{2}_{2}}{a_{3}}, \hspace{0.2 in} a_{4}= -\frac{a^{3}_{2}}{a^{2}_{3}}, \\
              b_{1}=\frac{b^{2}_{2}}{b_{3}}, \hspace{0.2 in} b_{4}= -\frac{b^{3}_{2}}{b^{2}_{3}}, \\
              c_{1}=\frac{c^{2}_{2}}{c_{3}}, \hspace{0.2 in} c_{4}= -\frac{c^{3}_{2}}{c^{2}_{3}}, \\
              d_{1}=\frac{d^{2}_{2}}{d_{3}}, \hspace{0.2 in} d_{4}= -\frac{d^{3}_{2}}{d^{2}_{3}}. 
              \label{wsp_high_symm2} 
           \end{gathered}
           \end{equation} 
           \end{itemize}             
          \end{enumerate}

          The parameters $\beta_{k}, (k = 1,...,5)$, occuring in (\ref{rozw_suma}) and 
          (\ref{argumenty1}), are, in this case, arbitrary constants. 
          
          We check now, whether the condition (\ref{warunek_h}) of existence of Legendre transformation
          (\ref{leg}) is satisfied for the ansatz (\ref{rozw_suma}) and for the three sets of the relations of 
          the coefficients (\ref{wsp_eq_symm}), (\ref{wsp_high_symm1}) and (\ref{wsp_high_symm2}).  
          It turns out that this condition is satisfied for this ansatz and for these above three sets of 
          relations of coefficients, if these below conditions are satisfied:
          
           \vspace{0.2 in} 
          
          \begin{itemize}
          
          \item for (\ref{wsp_eq_symm}) and for (\ref{wsp_high_symm2}), ($g_{j}, (j=1,...,4)$, are the 
          functions of the arguments, including the coefficients, 
          which satisfy correspondingly (\ref{wsp_eq_symm}) and (\ref{wsp_high_symm2}))

            \begin{equation}
            \begin{gathered}
          (b_{2}d_{3}-b_{3}d_{2})^{2}g''_{2}g''_{4}+(b_{2}c_{3}-b_{3}c_{2})^{2}g''_{2}g''_{3}+\\
					(a_{2}d_{3}-a_{3}d_{2})^{2}g''_{1}g''_{4}+
          (c_{2}a_{3}-c_{3}a_{2})^{2}g''_{1}g''_{3}+\\
					(a_{2}b_{3}-a_{3}b_{2})^{2}g''_{1}g''_{2}+(c_{2}d_{3}-c_{3}d_{2})^{2}g''_{3}g''_{4} \neq 
          0, 
          \label{w_rowne}
          \end{gathered}
          \end{equation}

          \item for (\ref{wsp_high_symm1})
          
            \begin{equation}
            \begin{gathered}
          \frac{( b_{2}^{4}a_{1}^{2}c_{1}^{2}d_{1}^{2}a_{2}^{2}-2a_{2}^{3}b_{1}c_{1}^{2}d_{1}^{2}b_{2}^{3}a_{1}
+a_{2}^{4}b_{1}^{2}c_{1}^{2}d_{1}^{2}b_{2}^{2}) g''_{1}g''_{2}}{a_{1}^{2}b_{1}^{2}c_{1}^{2}d_{1}^{2}}+\\
\frac{( -2a_{2}^{3}b_{1}^{2}c_{1}d_{1}^{2}c_{2}^{3}a_{{1}}+c_{2}^{4}a_{1}^{2}b_{1}^{2}d_{1}^{2}a_{2}^{2}
+a_{2}^{4}b_{1}^{2}c_{1}^{2}d_{1}^{2}c_{2}^{2}) g''_{1}g''_{3}}{a_{1}^{2}b_{1}^{2}c_{1}^{2}d_{1}^{2}}+\\
\frac{( -2a_{2}^{3}b_{1}^{2}c_{1}^{2}d_{1}d_{2}^{3}a_{1}+a_{2}^{4}b_{1}^{2}c_{1}^{2}d_{1}^{2}d_{2}^{2}
+d_{2}^{4}a_{1}^{2}b_{1}^{2}c_{1}^{2}a_{2}^{2}) g''_{1}g''_{4}}{a_{1}^{2}b_{1}^{2}c_{1}^{2}d_{1}^{2}}+\\
\frac{( c_{2}^{4}a_{1}^{2}b_{1}^{2}d_{1}^{2}b_{2}^{2}+b_{2}^{4}a_{1}^{2}c_{1}^{2}d_{1}^{2}c_{2}^{2}
-2b_{2}^{3}a_{1}^{2}c_{1}d_{1}^{2}c_{2}^{3}b_{1}) g''_{2}g''_{3}}{a_{1}^{2}b_{1}^{2}c_{1}^{2}d_{1}^{2}}+\\
\frac{(-2b_{2}^{3}a_{1}^{2}c_{1}^{2}d_{1}d_{2}^{3}b_{1}+b_{2}^{4}a_{1}^{2}c_{1}^{2}d_{1}^{2}d_{2}^{2}+d_{2}^{4}a_{1}^{2}b_{1}^{2}c_{1}^{2}b_{2}^{2}) g''_{2}g''_{4}}{a_{1}^{2}b_{1}^{2}c_{1}^{2}d_{1}^{2}}+\\
\frac{( d_{2}^{4}a_{1}^{2}b_{1}^{2}c_{1}^{2}c_{2}^{2}+c_{2}^{4}a_{1}^{2}b_{1}^{2}d_{1}^{2}d_{2}^{2}
-2c_{2}^{3}a_{1}^{2}b_{1}^{2}d_{1}d_{2}^{3}c_{1}) g''_{3}g''_{4}}{a_{1}^{2}b_{1}^{2}c_{1}^{2}d_{1}^{2}} \neq 0 
\label{w_wysokie},
          \end{gathered}
          \end{equation}  
          \end{itemize}
          
    where $g''_{j}, (j=1,2,3,4)$ denotes the second derivative of the function $g_{j}$ with respect to its 
    argument.
          
          Now, basing on the considerations included in \cite{malykh2004}, we make the 
          analysis of non-invariance of these above classes of exact solutions of second heavenly equation.
          Namely, we check now, whether $\Sigma_{1}, \Sigma_{2}, \Sigma_{3}, \Sigma_{4}$ are linearly 
          independent, i.e. the transformations from $x, r, t, z$ to $\Sigma_{j}, (j=1,2,3,4)$ are invertible. 
          The Jacobian matrices are the following:
          
          \begin{enumerate}
          
          \item for the case of equal symmetries (when the relations (\ref{wsp_eq_symm}) hold): \\
          \begin{equation}
          M = \left( \begin{array}{cccc}
          \frac{a_{2}(a_{2}+a_{3})}{a_{3}}g'_{1} &  a_{2}g'_{1} &  a_{3}g'_{1} & 
          -\frac{a_{2}(a_{2}+a_{3})^{2}}{a^{2}_{3}}g'_{1} \\
          \frac{b_{2}(b_{2}+b_{3})}{b_{3}}g'_{2} & b_{2}g'_{2} & b_{3}g'_{2} & 
          -\frac{b_{2}(b_{2}+b_{3})^{2}}{b^{2}_{3}}g'_{2} \\
      \frac{c_{2}(c_{2} + c_{3})}{c_{3}}g'_{3} & c_{2}g'_{3} & c_{3}g'_{3} & - \frac{c_{2}(c_{2} + 
      c_{3})^{2}}{c^{2}_{3}}g'_{3} \\
      \frac{d_{2}(d_{2} + d_{3})}{d_{3}}g'_{4} & d_{2}g'_{4} & d_{3}g'_{4} & - \frac{d_{2}(d_{2} + 
      d_{3})^{2}}{d^{2}_{3}}g'_{4}
      \end{array} \right ), \label{macierz1}
          \end{equation} 
          
         \item for the case of higher symmetry (when the relations (\ref{wsp_high_symm1}) hold) : \\          
          \begin{equation}
          M = \left( \begin{array}{cccc}
            a_{1}g'_{1} & a_{2}g'_{1} & \frac{a^{2}_{2}}{a_{1}}g'_{1} & -\frac{a^{2}_{1}}{a_{2}}g'_{1} \\
            b_{1}g'_{2} & b_{2}g'_{2} & \frac{b^{2}_{2}}{b_{1}}g'_{2} & -\frac{b^{2}_{1}}{b_{2}}g'_{2} \\
            c_{1}g'_{3} & c_{2}g'_{3} & \frac{c^{2}_{2}}{c_{1}}g'_{3} & -\frac{c^{2}_{1}}{c_{2}}g'_{3} \\
            d_{1}g'_{4} & d_{2}g'_{4} & \frac{d^{2}_{2}}{d_{1}}g'_{4} & -\frac{d^{2}_{1}}{d_{2}}g'_{4}
          \end{array} \right), \label{macierz2}
          \end{equation} 
          
          \item for the case of higher symmetry (when the relations (\ref{wsp_high_symm2}) hold) : \\  
           \begin{equation}
          M = \left( \begin{array}{cccc}
            \frac{a^{2}_{2}}{a_{3}} g'_{1} & a_{2}g'_{1} & a_{3}g'_{1} & -\frac{a^{3}_{2}}{a^{2}_{3}}g'_{1} \\
            \frac{b^{2}_{2}}{b_{3}} g'_{2} & b_{2}g'_{2} & b_{3}g'_{2} & -\frac{b^{3}_{2}}{b^{2}_{3}}g'_{2} \\
            \frac{c^{2}_{2}}{c_{3}}g'_{3} &  c_{2}g'_{3} & c_{3}g'_{3} & -\frac{c^{3}_{2}}{c^{2}_{3}}g'_{3} \\
            \frac{d^{2}_{2}}{d_{3}}g'_{4} &  d_{2}g'_{4} & d_{3}g'_{4} & -\frac{d^{3}_{2}}{d^{2}_{3}}g'_{4}
          \end{array} \right), \label{macierz3}
          \end{equation}

          \end{enumerate} 
          
          where $g'_{j}$ ($j = 1, 2, 3, 4$), is the first derivative of the function $g_{j}$ with respect to 
          its argument. We require non-vanishing of the Jacobians:

          \begin{itemize}
          \item - the case of equal symmetries
            \begin{equation}
            \begin{gathered}     
  \det{M} = 
\frac {g'_{1}g'_{2}g'_{3}g'_{4}}{a^{2}_{3}b^{2}_{3}d^{2}_{3}c^{2}_{3}}
\bigg(-a_{2}^{2}a_{{3}}b_{{2}}b_{3}^{2}c_{3}^{3}d_{2}^{3}+a_{2}^{2}a_{{3}}b_{{2}}b_{3}^{2}c_{2}^{3}d_{3}^{3} \\
-a_{2}^{2}a_{{3}}c_{{2}}c_{3}^{2}d_{3}^{3}b_{2}^{3}+
a_{2}^{2}a_{{3}}c_{{2}}c_{3}^{2}b_{3}^{3}d_{2}^{3}
-a_{2}^{2}a_{{3}}d_{{2}}d_{3}^{2}b_{3}^{3}c_{2}^{3}+\\
a_{2}^{2}a_{{3}}d_{{2}}d_{3}^{2}c_{3}^{3}b_{2}^{3}
+b_{2}^{2}b_{{3}}a_{{2}}a_{3}^{2}c_{3}^{3}d_{2}^{3}-b_{2}^{2}b_{{3}}a_{{2}}a_{3}^{2}c_{2}^{3}d_{3}^{3}\\
+b_{2}^{2}b_{{3}}c_{{2}}c_{3}^{2}d_{3}^{3}a_{2}^{3}-
b_{2}^{2}b_{{3}}c_{{2}}c_{3}^{2}a_{3}^{3}d_{2}^{3}
+b_{2}^{2}b_{{3}}d_{{2}}d_{3}^{2}a_{3}^{3}c_{2}^{3}-\\
b_{2}^{2}b_{{3}}d_{{2}}d_{3}^{2}c_{3}^{3}a_{2}^{3}
+c_{2}^{2}c_{{3}}a_{{2}}a_{3}^{2}d_{3}^{3}b_{2}^{3}-c_{2}^{2}c_{{3}}a_{{2}}a_{3}^{2}b_{3}^{3}d_{2}^{3}\\
-c_{2}^{2}c_{{3}}b_{{2}}b_{3}^{2}d_{3}^{3}a_{2}^{3}+
c_{2}^{2}c_{{3}}b_{{2}}b_{3}^{2}a_{3}^{3}d_{2}^{3}
-c_{2}^{2}c_{{3}}d_{{2}}d_{3}^{2}a_{3}^{3}b_{2}^{3}+\\
c_{2}^{2}c_{{3}}d_{{2}}d_{3}^{2}b_{3}^{3}a_{2}^{3}
+d_{2}^{2}d_{{3}}a_{{2}}a_{3}^{2}b_{3}^{3}c_{2}^{3}-d_{2}^{2}d_{{3}}a_{{2}}a_{3}^{2}c_{3}^{3}b_{2}^{3}\\
-d_{2}^{2}d_{{3}}b_{{2}}b_{3}^{2}a_{3}^{3}c_{2}^{3}+
d_{2}^{2}d_{{3}}b_{{2}}b_{3}^{2}c_{3}^{3}a_{2}^{3}
+d_{2}^{2}d_{{3}}c_{{2}}c_{3}^{2}a_{3}^{3}b_{2}^{3}-\\
d_{2}^{2}d_{{3}}c_{{2}}c_{3}^{2}b_{3}^{3}a_{2}^{3} \bigg) \neq 0, \label{wyzn_rown}
\end{gathered}
\end{equation}

        \item - the case of higher symmetry - subclass I
        
            \begin{equation}
            \begin{gathered}
        \det{M} = 
     \frac{g'_{1}g'_{2}g'_{3}g'_{4}}{c_{1}d_{2}c_{2}d_{1}b_{2}b_{1}a_{2}a_{1}} \bigg( -a_{1}^{2}a_{2}b_{2}^{2}b_{1}c_{2}^{3}d_{1}^{3}+\\
				a_{1}^{2}a_{{2}}b_{2}^{2}b_{{1}}c_{1}^{3}d_{2}^{3}
-a_{1}^{2}a_{2}c_{2}^{2}c_{1}d_{2}^{3}b_{1}^{3}+\\
a_{1}^{2}a_{{2}}c_{2}^{2}c_{{1}}b_{2}^{3}d_{1}^{3}-a_{1}^{2}a_{2}d_{2}^{2}d_{1}b_{2}^{3}c_{1}^{3}+\\
a_{1}^{2}a_{2}d_{2}^{2}d_{1}c_{2}^{3}b_{1}^{3}
+b_{1}^{2}b_{2}a_{2}^{2}a_{1}c_{2}^{3}d_{1}^{3}\\
-b_{1}^{2}b_{{2}}a_{2}^{2}a_{1}c_{1}^{3}d_{2}^{3}
+b_{1}^{2}b_{2}c_{2}^{2}c_{1}d_{2}^{3}{a_{{1}}}^{3}-\\
b_{1}^{2}b_{{2}}c_{2}^{2}c_{{1}}a_{2}^{3}d_{1}^{3}
+b_{1}^{2}b_{2}d_{2}^{2}d_{1}a_{2}^{3}c_{1}^{3}-\\
b_{1}^{2}b_{2}d_{2}^{2}d_{1}c_{2}^{3}a_{1}^{3}
+c_{1}^{2}c_{2}a_{2}^{2}a_{1}d_{2}^{3}b_{1}^{3}-\\
c_{1}^{2}c_{{2}}a_{2}^{2}a_{1}b_{2}^{3}d_{1}^{3}
-c_{1}^{2}c_{2}b_{2}^{2}b_{1}d_{2}^{3}a_{1}^{3}+\\
c_{1}^{2}c_{{2}}b_{2}^{2}b_{1}a_{2}^{3}d_{1}^{3}
-c_{1}^{2}c_{2}d_{2}^{2}d_{1}a_{2}^{3}b_{1}^{3}+\\
c_{1}^{2}c_{2}d_{2}^{2}d_{1}b_{2}^{3}a_{1}^{3}
+d_{1}^{2}d_{2}a_{2}^{2}a_{1}b_{2}^{3}c_{1}^{3}-\\
d_{1}^{2}d_{{2}}a_{2}^{2}a_{1}c_{2}^{3}b_{1}^{3}
-d_{1}^{2}d_{2}b_{2}^{2}b_{1}a_{2}^{3}c_{1}^{3}+\\
d_{1}^{2}d_{{2}}b_{2}^{2}b_{1}c_{2}^{3}a_{1}^{3}
+d_{1}^{2}d_{2}c_{2}^{2}c_{1}a_{2}^{3}b_{1}^{3}-\\
d_{1}^{2}d_{2}c_{2}^{2}c_{1}b_{2}^{3}a_{1}^{3} \bigg) \neq 0,   \label{wyzn_wys1}
  \end{gathered}
  \end{equation}

    \item - the case of higher symmetry - subclass II
    
   \begin{equation}
   \begin{gathered}   
  \det{M} = 
  \frac{g'_{1}g'_{2}g'_{3}g'_{4}}{a^{2}_{3}b^{2}_{3}d^{2}_{3}c^{2}_{3}} \bigg(
  -a^{2}_{2}a_{3}b_{2}b^{2}_{3}c^{3}_{3}d^{3}_{2}+a^{2}_{2}a_{3}b_{2}b^{2}_{3}c^{3}_{2}d^{3}_{3}
-\\
a^{2}_{2}a_{3}c_{2}c^{2}_{3}d^{3}_{3}b^{3}_{2}+a^{2}_{2}a_{3}c_{2}c^{2}_{3}b^{3}_{3}d^{3}_{2}\\
-a^{2}_{2}a_{3}d_{2}d^{2}_{3}b^{3}_{3}c^{3}_{2}+a^{2}_{2}a_{3}d_{2}d^{2}_{3}c^{3}_{3}b^{3}_{2}\\
+b^{2}_{2}b_{{3}}a_{2}a^{2}_{3}c^{3}_{3}d^{3}_{2}-b^{2}_{2}b_{3}a_{2}a^{2}_{3}c^{3}_{2}d^{3}_{3}\\
+b^{2}_{2}b_{3}c_{2}c^{2}_{3}d^{3}_{3}a^{3}_{2}-b^{2}_{2}b_{3}c_{2}c^{2}_{3}a^{3}_{3}d^{3}_{2}\\
+b^{2}_{2}b_{3}d_{2}d^{2}_{3}a^{3}_{3}c^{3}_{2}-b^{2}_{2}b_{3}d_{2}d^{2}_{3}c^{3}_{3}a^{3}_{2}\\
+c^{2}_{2}c_{3}a_{2}a^{2}_{3}d^{3}_{3}b^{3}_{2}-c^{2}_{2}c_{3}a_{2}a^{2}_{3}b^{3}_{3}d^{3}_{2}\\
-c^{2}_{2}c_{3}b_{2}b^{2}_{3}d^{3}_{3}a^{3}_{2}+c^{2}_{2}c_{3}b_{2}b^{2}_{3}a^{3}_{3}d^{3}_{2}\\
-c^{2}_{2}c_{3}d_{2}d^{2}_{3}a^{3}_{3}b^{3}_{2}+c^{2}_{2}c_{3}d_{2}d^{2}_{3}b^{3}_{3}a^{3}_{2}\\
+d^{2}_{2}d_{3}a_{2}a^{2}_{3}b^{3}_{3}c^{3}_{2}-d^{2}_{2}d_{3}a_{2}a^{2}_{3}c^{3}_{3}b^{3}_{2}\\
-d^{2}_{2}d_{3}b_{2}b^{2}_{3}a^{3}_{3}c^{3}_{2}+d^{2}_{2}d_{3}b_{2}b^{2}_{3}c^{3}_{3}a^{3}_{2}\\
+d^{2}_{2}d_{3}c_{2}c^{2}_{3}a^{3}_{3}b^{3}_{2}-d^{2}_{2}d_{3}c_{2}c^{2}_{3}b^{3}_{3}a^{3}_{2} \bigg) \neq 0.  \label{wyzn_wys2}
  \end{gathered}
  \end{equation} 
  
          \end{itemize}
          
          We can now repeat the reasonings: from the \cite{malykh2004}, that the Killing equation cannot be satisfied identically for any solution of Legendre transformed second heavenly equation of Pleba\'{n}ski (\ref{heavenly_leg}) and from previous subsubsection.\\              
In these above three cases: $\det{(M}) \neq 0$, when the corresponding polynomials, included in (\ref{wyzn_rown}), (\ref{wyzn_wys1}), (\ref{wyzn_wys2}), do not possess zeroes and $g'_{1}g'_{2}g'_{3}g'_{4} \neq 0$ . So, after assumption that $a_{3} b_{3} c_{3} d_{3} \neq 0$ (the case of equal symmetries) and  $c_{1}d_{2}c_{2}d_{1}b_{2}b_{1}a_{2}a_{1} \neq 0$, $a_{3}b_{3}d_{3}c_{3} \neq 0$ (the case of higher symmetries), we may say that  $\Sigma_{1} = a_{1} x + a_{2} r + a_{3} t + a_{4} z + \beta_{2},...,
    \Sigma_{4} = d_{1} x + d_{2} r + d_{3} t + d_{4} z + \beta_{5} $ (where the coefficients satisfy (\ref{wsp_eq_symm}), (\ref{wsp_high_symm1}) and (\ref{wsp_high_symm2}), correspondingly),  are linearly independent and the transformations from $x, r, t, z$ to $\Sigma_{j}, (j=1, 2, 3, 4)$, are invertible. Then, we can express $x, r, t, z$ by $\Sigma_{j}, j=1,2,3,4$, so we can choose $\Sigma_{j}, j=1,2,3,4$, as new independent variables in Killing equation (derived in \cite{malykh2004}). After inserting each of above classes of solutions into this equation, we obtain a relation of the form:

          \begin{gather}
             F_{4}(\Sigma_{1}, \Sigma_{2}, \Sigma_{3}, \Sigma_{4}) = 0. \label{po_wybraniu}
          \end{gather} 
      
     These above classes of exact solutions have been obtained, by solving systems of second-order linear equations together with Legendre-transformed second heavenly equation (\ref{heavenly_leg}), they are determined up to arbitrary constants, because we may choose each of the function $g_{j}=f(\Sigma_{j})$, as $f(\delta + \varepsilon \Sigma_{j})$, where $\delta, \varepsilon$ are arbitrary constants.
      Let us {\em \bf fix} now the functions $g_{k}, (k=1,...,4)$ in found classes of solutions,
      but such that the conditions (\ref{wyzn_rown}) (together with (\ref{w_rowne})), (\ref{wyzn_wys1}) (together with (\ref{w_wysokie})) and (\ref{wyzn_wys2}) (together with (\ref{w_rowne})), are satisfied, correspondingly. For example, if we choose $g_{j}=\exp{(\Sigma_{j})}$ in the above found classes of solutions,  we obtain solutions similar to the solutions given either by (\ref{rozw_malykh2004}) and (\ref{w_rownsym}) or by (\ref{rozw_malykh2004}) and (\ref{w_wyssym}), if $n=4$. 
       So, having no functional arbitrariness, the solutions, belonging to these above classes with fixed functions $g_{k}, (k=1,...,4)$, cannot satisfy in addition the first-order equation (Killing equation derived in \cite{malykh2004}).
      Hence, this equation cannot be tautology for the solutions belonging to these above classes, for the Legendre-transformed potential $\Pi(x,r,t,z)$, satisfying corresponding equation (derived in \cite{malykh2004}), and for suitable choice of the functions  $q,c,e,\rho,\sigma,\psi$ and the constants $a,k$.\\          
        
      Thus, we have showed that the metric (\ref{metryka_leg}) with $\vartheta$, being a solution, belonging to the classes, given by (\ref{rozw_suma}) and $\Sigma_{1} = a_{1} x + a_{2} r + a_{3} t + a_{4} z + \beta_{2},...,\Sigma_{4} = d_{1} x + d_{2} r + d_{3} t + d_{4} z + \beta_{5} $ (where in the case of the system (\ref{eq_symm1})-(\ref{eq_symm3}), the coefficients satisfy the relations (\ref{wsp_eq_symm}) and in the case of the system (\ref{high_symm1})-(\ref{high_symm3}), the 
          coefficients satisfy the relations (\ref{wsp_high_symm1}) or (\ref{wsp_high_symm2})) and the 
          functions $g_{j}, (j=1,...,4)$ are fixed, does not possess Killing vector. \\
           
      In order to obtain some extensions of classes of the solutions, given by (\ref{rozw_suma}) and (\ref{wsp_eq_symm}), (\ref{wsp_high_symm1}),  (\ref{wsp_high_symm2}) - the series of $n$ (we assume as yet that $n$ is a finite number) functions $g_{i}$, it is convenient to write the ansatz down in the convention applied in the formula (\ref{rozw_malykh2004}) from \cite{malykh2004}. Namely, the argument in each function $g_{j}$, is now: $\alpha_{j} x + \gamma_{j} r + \zeta_{j} t + \lambda_{j} z + \beta_{j}$. Hence, our ansatz has the form:
      
      \begin{gather} 
       \vartheta(x,r,t,z)=\sum^{n}_{j=1} g_{j}(\Sigma_{j}), \label{rozw_suma_i} 
      \end{gather}
      
      where $g_{j}$ are {\em arbitrary} holomorphic functions of:
       
       \begin{gather}
       \Sigma_{j} = \alpha_{j} x + \gamma_{j} r + \zeta_{j} t + \lambda_{j} z + \beta_{j}. \label{argumenty_i} 
       \end{gather}
      
     Obviously, now the notation of the coefficients changes. We give here this change for $j=1,...,4$:
     
     \begin{equation}
     \begin{gathered}
     a_{1}=\alpha_{1}, \hspace{0.1 in} a_{2}=\gamma_{1}, \hspace{0.1 in} a_{3}=\zeta_{1}, \hspace{0.1 in} 
     a_{4} = \lambda_{1}, \\
     b_{1}=\alpha_{2}, \hspace{0.1 in} b_{2}=\gamma_{2}, \hspace{0.1 in} b_{3}=\zeta_{2}, \hspace{0.1 in} 
     b_{4} = \lambda_{2}, \\
     c_{1}=\alpha_{3}, \hspace{0.1 in} c_{2}=\gamma_{3}, \hspace{0.1 in} c_{3}=\zeta_{3}, \hspace{0.1 in} 
     c_{4} = \lambda_{3}, \\
     d_{1}=\alpha_{4}, \hspace{0.1 in} d_{2}=\gamma_{4}, \hspace{0.1 in} d_{3}=\zeta_{4}, \hspace{0.1 in} 
     d_{4} = \lambda_{4}. \label{zmiana_oznaczen} 
     \end{gathered}
     \end{equation}

      Then, (\ref{rozw_suma_i}) is some class of solutions of the systems (\ref{eq_symm1})-(\ref{eq_symm3}) 
      and (\ref{high_symm1})-(\ref{high_symm3}) (correspondingly) and in consequence, of second heavenly 
      equation (\ref{heavenly_leg}), if the coefficients satisfy the 
      following relations: 
      
      \begin{enumerate}
       \item for the case of equal symmetries: \\
       
       \begin{equation} 
    \alpha_{j}=\frac{\gamma_{j}(\gamma_{j} + \zeta_{j})}{\zeta_{j}}, \hspace{0.1 in} 
    \lambda_{j}=-\frac{\gamma_{j}(\gamma_{j}+\zeta_{j})^{2}}{\zeta^{2}_{j}},
    \label{wsp_suma_rowne}
       \end{equation}
       
       \item for the case of higher symmetry we found two subclasses: \\
      
         \begin{itemize}
         \item I subclass
         
         \begin{equation} 
        \zeta_{j}=\frac{\gamma^{2}_{j}}{\alpha_{j}}, \hspace{0.1 in} 
        \lambda_{j}=-\frac{\alpha^{2}_{j}}{\gamma_{j}}, \label{wsp_suma_wys_1}
       \end{equation}
      
       \item II subclass
       
        \begin{equation} 
    \alpha_{j}=\frac{\gamma^{2}_{j}}{\zeta_{j}}, \hspace{0.1 in} 
    \lambda_{j}=-\frac{\gamma^{3}_{j}}{\zeta^{2}_{j}}. \label{wsp_suma_wys_2}
       \end{equation}
       \end{itemize}
      \end{enumerate}  
      
      In both cases: of equal symmetries and higher symmetry, $\beta_{j}$ are arbitrary constants.
			It turns out that in the case of higher symmetry, the class of the solutions of second heavenly equation 
			transformed by Legendre transformation, is given by functional series, which appear in (\ref{rozw_suma_i}), 
			and this series can be infinite.

      However, we have here the functional series (\ref{rozw_suma_i}). So, now we need to apply some properties of the 
      functional series, 
			\cite{korn}. Namely, from the requirement of differentiability of (\ref{rozw_suma_i}), we have 
      the requirement 
      that it needs to be uniformly convergent. Also, from the requirement of differentiability of 
      (\ref{rozw_suma_i}), we see that the corresponding series (for the coefficients satisfying the 
      relations (\ref{wsp_suma_rowne})):  
      
       \begin{equation}
       \begin{gathered}
      \sum^{n}_{j=1} \frac{\partial}{\partial 
      x} g_{j} \hspace{0.08 in}, \hspace{0.05 in} ... \hspace{0.05 in}, \hspace{0.08 in} 
      \sum^{n}_{j=1} \frac{\partial}{\partial z} g_{j}, \label{rozn_szereg}
      \end{gathered}
      \end{equation}
      
      need to be uniformly convergent.
      
      Further, from the requirement of differentiability of the series (\ref{rozn_szereg}), the consecutive 
      series, including the terms obtained by computing the derivatives in (\ref{rozn_szereg}) (for the 
      coefficients satisfying the relations (\ref{wsp_suma_rowne})):
      
      \begin{gather} 
      \sum^{n}_{j=1} \frac{\partial}{\partial x} 
      \bigg( g'_{j}\frac{\gamma_{j}(\gamma_{j}+\zeta_{j})}{\zeta_{j}} \bigg)  \hspace{0.08 in} ,  
      \hspace{0.08 in} \sum^{n}_{j=1} \frac{\partial}{\partial t} 
      \bigg(g'_{j}\frac{\gamma_{j}(\gamma_{j}+\zeta_{j})}{\zeta_{j}}\bigg), \\
      \hspace{0.08 in} etc. \hspace{0.08 in} etc. \label{drugie_p_szer_r}
      \end{gather}

      need to be uniformly convergent, too, that the system (\ref{eq_symm1})-(\ref{eq_symm3}), for 
      (\ref{rozw_suma_i}), when the relations (\ref{wsp_suma_rowne}), is satisfied.

      One can also check that the ansatz (\ref{rozw_suma_i}) with the parameters satisfying (\ref{wsp_suma_rowne}), satisfies 
			the equation (\ref{heavenly_leg}), too.

       Analogically, in the case of higher symmetry or of the system (\ref{high_symm1})-(\ref{high_symm3}), 
       the series (\ref{rozw_suma_i}) needs also to be uniformly convergent. 
       Also, (for the case of the relations (\ref{wsp_suma_wys_1}) and (\ref{wsp_suma_wys_2}))
      from the requirement of differentiability of (\ref{rozw_suma_i}), we see that 
      the corresponding series: 
       
      \begin{equation}
      \begin{gathered}
      \sum^{n}_{j=1} \frac{\partial}{\partial x} g_{j} 
      \hspace{0.08 in}, \hspace{0.05 in} ... \hspace{0.05 in}, \hspace{0.08 in} 
      \sum^{n}_{j=1} \frac{\partial}{\partial z} g_{j}, \label{rozn_szereg2}
      \end{gathered}
      \end{equation}
      
      need to be uniformly convergent.
      
      Further, from the requirement of differentiability of the series (\ref{rozn_szereg2}), the consecutive 
      series, including the terms obtained by computing the derivatives in (\ref{rozn_szereg2}):
      
      \begin{itemize}
      \item when the relations (\ref{wsp_suma_wys_1}) hold:
      
      \begin{gather} 
      \sum^{n}_{j=1} \frac{\partial}{\partial x} 
      \bigg(g'_{j} \alpha_{j}\bigg) \hspace{0.08 in}, \hspace{0.08 in} \sum^{n}_{j=1} 
      \frac{\partial}{\partial x} \bigg(g'_{j} \gamma_{j} \bigg),
      \hspace{0.08 in} etc. \hspace{0.08 in} etc.
      \end{gather}
      
      \item when the relations (\ref{wsp_suma_wys_2}) hold:
      
      \begin{gather} 
      \sum^{n}_{j=1} \frac{\partial}{\partial x} 
      \bigg(g'_{j} \frac{\gamma^{2}_{j}}{\zeta_{j}}\bigg) \hspace{0.08 in}, \hspace{0.08 in} 
      \sum^{n}_{j=1} \frac{\partial}{\partial x} 
      \bigg(g'_{j} \gamma_{j}\bigg),
      \hspace{0.08 in} etc. \hspace{0.08 in} etc.
      \end{gather}
      
      \end{itemize} 
      
      need to be uniformly convergent, too. One can check that the system (\ref{high_symm1})-(\ref{high_symm3}), is satisfied by 
      (\ref{rozw_suma_i}), when the coefficients satisfy the relations (\ref{wsp_suma_wys_1}) and similar situation is 
      in the case, when the coefficients satisfy the relations (\ref{wsp_suma_wys_2}).
      
          As we see, the ansatz (\ref{rozw_suma_i}), with the relations (\ref{argumenty_i}), (\ref{wsp_suma_wys_1}), is some direct generalization of the solution given by (\ref{rozw_malykh2004}) and (\ref{w_wyssym}), found in \cite{malykh2004}. 
      We check now, whether the condition (\ref{warunek_h}) is satisfied for these three classes of solutions given by the ansatz (\ref{rozw_suma_i}) and (\ref{wsp_suma_rowne}), (\ref{wsp_suma_wys_1}), (\ref{wsp_suma_wys_2}), correspondingly. It turns out that it is satisfied for these three classes, if the following relations hold:
      
      \begin{enumerate}
      \item - for (\ref{wsp_suma_rowne})

      \begin{gather}
      \bigg(\sum^{n}_{j=1} g''_{j} \zeta^{2}_{j} \bigg) \bigg(\sum^{n}_{j=1} g''_{j} \gamma^{2}_{j} \bigg) - 
      \bigg(\sum^{n}_{j=1} g''_{j} \gamma_{j} \zeta_{j} \bigg)^{2} \neq 0, \label{war_leg_h_i_1}
      \end{gather}
      
      \item - for (\ref{wsp_suma_wys_1})
      
      \begin{gather}
      \bigg (\sum^{n}_{j=1} \frac{g''_{j} \gamma^{4}_{j}}{\alpha^{2}_{j}} \bigg) \bigg(\sum^{n}_{j=1} g''_{j} 
      \gamma^{2}_{j} \bigg) - 
      \bigg(\sum^{n}_{j=1} \frac{g''_{j} \gamma^{3}_{j}}{\alpha_{j}} \bigg)^{2} \neq 0, \label{war_leg_h_i_2}
      \end{gather}

      \item - for (\ref{wsp_suma_wys_2})

      \begin{gather}
      \bigg(\sum^{n}_{j=1} g''_{j} \zeta^{2}_{j} \bigg) \bigg(\sum^{n}_{j=1} g''_{j} \gamma^{2}_{j} \bigg) - 
      \bigg(\sum^{n}_{j=1} g''_{j} 
      \gamma_{j} \zeta_{j} \bigg)^{2} \neq 0. \label{war_leg_h_i_3}
      \end{gather}
      \end{enumerate}

       The conditions (\ref{wyzn_rown}), (\ref{wyzn_wys1}), (\ref{wyzn_wys2}) are satisfied also for 
       (\ref{rozw_suma_i}) - (\ref{argumenty_i}) and (\ref{wsp_suma_rowne}), (\ref{wsp_suma_wys_1}), (\ref{wsp_suma_wys_2}), 
      correspondingly, but, of course, the notation for the coefficients $a_{j}, b_{j}, c_{j}, d_{j}, (j=1,...,4)$, 
      changes according to (\ref{zmiana_oznaczen}). \\
      Hence, now we may repeat the similar reasonings (included in \cite{malykh2004}), as previously, and we may say that if $n \geq 4$, then  $\Sigma_{j} = \alpha_{j} x + \gamma_{j} r + \zeta_{j} t + \lambda_{j} z + \beta_{j}, (j=1,2,3,4)$, are linearly independent and the transformations from $x, r, t, z$ to $\Sigma_{j}, (j=1, 2, 3, 4)$ are invertible, when the conditions: (\ref{wyzn_rown}) - for the equal symmetries and (\ref{wyzn_wys1}), (\ref{wyzn_wys2}) - for the higher symmetry, are satisfied (after taking into consideration the relations (\ref{zmiana_oznaczen})). Hence, we can express $x, r, t, z$ by $\Sigma_{i}$ and the same for $\Sigma_{5},...,\Sigma_{n}$, so we can choose $\Sigma_{j}, j=1,2,3,4$, as new independent variables in Killing equation  (derived in \cite{malykh2004}). Hence, after inserting any solution belonging to each of the classes, given either by (\ref{rozw_suma_i}), (\ref{argumenty_i}), (\ref{wsp_suma_rowne}) or by (\ref{rozw_suma_i}), (\ref{argumenty_i}), (\ref{wsp_suma_wys_1}) or 
      by (\ref{rozw_suma_i}), (\ref{argumenty_i}), (\ref{wsp_suma_wys_2}), correspondingly, into the Killing equation  mentioned above, 
      this equation will possess the form like (\ref{po_wybraniu}). 
      
      The solutions belonging to such classes, obtained by solving the systems of linear equations together with Legendre-transformed second heavenly equation  (\ref{heavenly_leg}), are determined up to arbitrary constants, because we may choose each of the function $g_{j}=f(\Sigma_{j})$, as $f(\delta + \varepsilon \Sigma_{j})$, where $\delta, \varepsilon$ are arbitrary constants. Now, let us {\em \bf fix} the functions $g_{j}, (j=1,...,n)$ in (\ref{rozw_suma_i}) for each of obtained classes, but such that the conditions (\ref{wyzn_rown})
      (together with (\ref{war_leg_h_i_1})), (\ref{wyzn_wys1}) (together with (\ref{war_leg_h_i_2})) and (\ref{wyzn_wys2}) (together with (\ref{war_leg_h_i_3})) are satisfied, correspondingly (of course, the notation for the coefficients $a_{j}, b_{j}, c_{j}, d_{j}, (j=1,...,4)$, changes according to (\ref{zmiana_oznaczen}) and so, the relations: $\zeta_{1}\zeta_{2}\zeta_{3}\zeta_{4} \neq 0$, $\alpha_{1}\gamma_{1}\alpha_{2}\gamma_{2}\alpha_{3}\gamma_{3}\alpha_{4}\gamma_{4} \neq 0$ and $\zeta_{1}\zeta_{2}\zeta_{3}\zeta_{4} \neq 0$,
      need to hold, correspondingly). For example, if we choose $g_{j}=\exp{(\Sigma_{j})}$ in (\ref{rozw_suma_i}), then we obtain form similar to the form of the ansatz (\ref{rozw_malykh2004}). 
      So, after fixing functions $g_{j}, (j=1,...,n)$, these solutions, having no functional arbitrariness,  cannot solve in addition, the first-order Killing equation (derived in \cite{malykh2004}).\\
            Hence, this equation cannot be tautology for the solutions belonging to these above three classes, for the Legendre-transformed potential $\Pi(x,r,t,z)$, satisfying appropriate equation (derived in \cite{malykh2004}), for suitable choice of the functions $q,c,e,\rho,\sigma,\psi$ and the constants $a,k$. 
            
      Thus, we have proved the following theorem:

        \begin{thm}
        The metric (\ref{metryka_leg}) with $\vartheta$, being exact solution of (\ref{heavenly_leg}), 
        belonging to any class, defined by (\ref{rozw_suma_i}) (where $g_{i}$ are the 
        functions of (\ref{argumenty_i})), when $n$ is an arbitrary natural number, and by the relations: 
        \begin{enumerate}
        \item (\ref{wsp_suma_rowne}) - class I \hspace{0.05 in} (the 
        case of equal symmetries), 
        \item (\ref{wsp_suma_wys_1}) - subclass I \hspace{0.05 in} (the case I of higher symmetry),
        \item (\ref{wsp_suma_wys_2}) - subclass II \hspace{0.05 in} (the case II of higher symmetry),  
        \end{enumerate}
        
        where the functions $g_{j}, (j=1,...,n)$ are fixed, does not possess Killing vector, when $n \geq 4$ and the conditions:   (\ref{wyzn_rown}),  $\zeta_{1}\zeta_{2}\zeta_{3}\zeta_{4} \neq 0$, (\ref{war_leg_h_i_1}), $n$ is finite  - for the class I (the case of equal symmetries), (\ref{wyzn_wys1}),  $\alpha_{1}\gamma_{1}\alpha_{2}\gamma_{2}\alpha_{3}\gamma_{3}\alpha_{4}\gamma_{4} \neq 0$, (\ref{war_leg_h_i_2}) - for the subclass I (the case of higher symmetry) and (\ref{wyzn_wys2}), $\zeta_{1}\zeta_{2}\zeta_{3}\zeta_{4} \neq 0$, (\ref{war_leg_h_i_3}) - for the subclass II (the case of higher symmetry), after taking into consideration the relations (\ref{zmiana_oznaczen}), are satisfied. In the case of higher symmetry, the class of the solutions is given by an infinite functional series. \\
				\end{thm}

        In all these case, the series  (\ref{rozw_suma_i}) some other corresponding series need to be uniformly convergent.				

        In the case of the higher symmetries, one can check by using Maple Waterloo Software that also 
				the ansatz $\vartheta(x,r,t,z)=\sum^{\infty}_{j=1} g_{j}(\Sigma_{j})$, (where $g_{j}$ are arbitrary 
				holomorphic functions of $\Sigma_{j} = \alpha_{j} x + \gamma_{j} r + \zeta_{j} t + 
				\lambda_{j} z + \beta_{j}$), gives the class of the solutions of the system (\ref{high_symm1})-(\ref{high_symm3})
				and also of (\ref{heavenly_leg}), when the relations (\ref{wsp_suma_wys_1}) and (\ref{wsp_suma_wys_2}) hold. One can also check that the condition (\ref{warunek_h}) is satisfied in the case of this above ansatz
				and the relations (\ref{wsp_suma_wys_1}) and (\ref{wsp_suma_wys_2}). 
				We can formulate the following theorem. 
				
		\begin{thm}
	   The second heavenly equation (\ref{heavenly_leg}) (obtained by a Legendre transformation of (\ref{heavenly})) and 
		 original version of the second heavenly equation (\ref{heavenly}), {\em{not transformed by the Legendre transformation}}, 
		 possess the class  of solutions of the form of the functional series 
				
		\begin{gather}
                u(x^{\mu})=\beta_{1} + \sum^{n}_{j = 1} g_{j}(\Sigma_{j}), \label{rozw_suma_niesk} 
                 \end{gather}
 
         where $u=v$ for the case of (\ref{heavenly}) and $u=\vartheta$ for the case of (\ref{heavenly_leg}), 
				$g_{j}$ are some arbitrary holomorphic functions of the arguments $\Sigma_{m} = a^{(m)}_{\nu} x^{\nu} + \beta_{m+1}$,
				 ($n$ can be any natural number, so the series $\sum^{n}_{j = 1} g_{j}(\Sigma_{j})$ can be finite or infinite), 
				$x^{\nu}$ are proper independent variables, the series $\sum^{n}_{j = 1} g_{j}(\Sigma_{j})$ and its corresponding
				derivatives, are uniformly convergent, $a^{(m)}_{\nu}$ are some constants 
				satisfying some relations following from satisfying of the system of algebraic equations, following from applying of the 
				decomposition method to the versions of second heavenly equation: (\ref{heavenly_leg}) and (\ref{heavenly}), and $\beta_{m+1}$ are arbitrary constants. 
				\end{thm}
						
				\begin{proof}
              This is sufficient to prove that for any $n > 0$ ($n \in {1, 2, 3,...}$), 
							of course, afterproper changing of the independent variables in the ansatz 
							(\ref{rozw_suma_niesk}), 
              one can decompose the equation obtained after inserting this ansatz into
							(\ref{heavenly_leg}) and (\ref{heavenly}), 
              according to the idea of the decomposition method \cite{stepien2010}, and next, that according to this method, 
              one can obtain a system of algebraic equations. The solutions of this system establish the relations between 
              the coefficients, which occur in (\ref{rozw_suma_niesk}). If these relations are satisfied, 
							then the ansatz (\ref{rozw_suma_niesk}) gives some class of the solutions 
							of the equations (\ref{heavenly_leg}) and (\ref{heavenly}) 
              We prove this theorem by mathematical induction \cite{Rasiowa1973}.

			\begin{enumerate}
			\item we check, whether this Theorem holds for $n = 1$, after substituting the ansatz 
			(\ref{rozw_suma_niesk}) into (\ref{heavenly_leg}) and (\ref{heavenly}), we obtain the equation
							
			\begin{equation}
			g''_{1} a^{(1)}_{1} a^{(1)}_{3} + g''_{1} a^{(1)}_{2} a^{(1)}_{4} = 0,
		        \end{equation}
									
			so: $g''_{1} (a^{(1)}_{1} a^{(1)}_{3} + a^{(1)}_{2} a^{(1)}_{4}) = 0$ 
		        and it suffices to find the solution of the algebraic equation
			$a^{(1)}_{1} a^{(1)}_{3} + a^{(1)}_{2} a^{(1)}_{4} =  0$. 
									
		       \item we prove that if this Theorem holds for $n = k$, then this holds also for $n = k+1$		
							
			After substituting the ansatz $u(x^{\mu})=\beta_{1} + \sum^{n=k}_{j = 1} g_{j}(\Sigma_{j})$, into
			(\ref{heavenly_leg}) and (\ref{heavenly}), 
			and collecting the algebraic terms by $g''_{i} g''_{j}$, $g''_{i}$, one obtains the following differential equation (it follows from the fact that the equations (\ref{heavenly_leg}) 
			and (\ref{heavenly}) satisfy the assumptions of the decomposition method)
							
			\begin{gather}
		        g''^{2}_{1} X_{1} + g''_{1} g''_{2} X_{2} + g''_{1} g''_{3} X_{3} +...+ g''_{1} g''_{m} X_{m} +  \\
			      g''^{2}_{2} X_{2}  +...+ g''_{k-1} g''_{k} X_{k-1} + (g''_{k})^{2} X_{k} = 0,
		        \end{gather}
							
			      where $X_{i}, (i = 1,...,k)$ are some polynomials including the constants $a^{(k)}_{\nu}$. If we demand vanishing of these  
		        polynomials, we obtain some system of algebraic equations. Its solutions establish the relations between $a^{(n)}_{\nu}$.
		        Next, after substituting the ansatz $u(x^{\mu})=\beta_{1} + \sum^{n=k+1}_{n = 1} g_{j}(\Sigma_{j})$, into 
						(\ref{heavenly_leg}) and (\ref{heavenly}), 
		        and collecting the algebraic terms by $g''_{i} g''_{j}$, $g''_{i}$, one obtains the following differential equation (it also follows from the fact that the equations (\ref{heavenly_leg}) 
			and (\ref{heavenly}) satisfy the assumptions of the decomposition method)
							
							\begin{gather}
							g''^{2}_{1} X_{1} + g''_{1} g''_{2} X_{2} + g''_{1} g''_{3} X_{3} +...+ g''_{1} g''_{m} X_{m} +  \\
							g''^{2}_{2} X_{2}  +...+ g''_{k} g''_{k+1} X_{k} + (g''_{k+1})^{2} X_{k+1} = 0,
							\end{gather}
							
							where $X_{i}, i = 1,...,k+1$ are some polynomials including the constants $a^{(k+1)}_{\nu}$. If we demand vanishing of 
							these polynomials, we obtain again some system of algebraic equations. Similarly, as previously,
							its solutions establish the relations between $a^{(n)}_{\nu}$.
					
							\end{enumerate} 
							
							  q.e.d.
							
							\end{proof}

        In all these above considerations, the conditions: $g''_{j} \neq 0$ need to be satisfied.

 \subsection{Classes of exact solutions of mixed heavenly equation}

 In order to find more general solutions of mixed heavenly equation (\ref{mixed}), instead of solving equation (\ref{leg_mixed}), obtained from (\ref{mixed}) by Legendre transformation (\ref{legendre_tr}) , we apply decomposition method to the system (\ref{heav_lin1})-(\ref{heav_lin3}).\\

 Similarly to the case of second heavenly equation, we look for the class of solutions, given by the ansatz (\ref{rozw_suma}), but here: $u \equiv w$,  $x^{1} = \eta, x^{2} = \xi, x^{3} = q, x^{4} = y$, and $g_{j}$ ($j = 1, 2, 3, 4$), are  {\em arbitrary} functions of their arguments.     
      The ansatz (\ref{rozw_suma}) presents a class of solutions of the system 
      (\ref{heav_lin1})-(\ref{heav_lin3}), when the following relations are satisfied:

     \begin{equation}
     \begin{gathered}
      a_{3}=\frac{a^{2}_{1}+a^{2}_{2}}{a_{2}}, \hspace{0.2 in} a_{4}=-\frac{a^{2}_{1}a_{2}+a^{3}_{2}-a^{3}_{1}-a_{1}a^{2}_{2}}{a^{2}_{2}}, \\
      b_{3}=\frac{b^{2}_{1}+b^{2}_{2}}{b_{2}}, \hspace{0.2 in} b_{4}=-\frac{b^{2}_{1}b_{2}+b^{3}_{2}-b^{3}_{1}-b_{1}b^{2}_{2}}{b^{2}_{2}}, \\
   c_{3}=\frac{c^{2}_{1}+c^{2}_{2}}{c_{2}}, \hspace{0.2 in} c_{4}=-\frac{c^{2}_{1}c_{2}+c^{3}_{2}-c^{3}_{1}-c_{1}c^{2}_{2}}{c^{2}_{2}}, \\
   d_{3}=\frac{d^{2}_{1}+d^{2}_{2}}{d_{2}}, \hspace{0.2 in} d_{4}=-\frac{d^{2}_{1}d_{2}+d^{3}_{2}-d^{3}_{1}-d_{1}d^{2}_{2}}{d^{2}_{2}}.  
  \label{wsp_mixed}  
      \end{gathered}
      \end{equation}

     It turns out that the condition (\ref{warunek_mix}) is satisfied for the above class of solutions, when
     relation (\ref{war_leg_mix}) holds (see Appendix A).\\          
     So, now we check, whether the class (\ref{rozw_suma}), is a class of non-invariant solutions or it    
     depends on four independent combinations of the variables $\eta, \xi, q, y$, for these relations between the coefficients (\ref{wsp_mixed}). Analogously, as in the case of second heavenly euqation, we write down the Jacobian matrix:

     \begin{equation}
          M = \left( \begin{array}{cccc}
       a_{1}g'_{1} & a_{2}g'_{1} & \frac{a^{2}_{1}+a^{2}_{2}}{a_{2}}g'_{1} & -\frac{a^{2}_{1}a_{2}+a^{3}_{2}-a^{3}_{1}-a_{1}a^{2}_{2}}{a^{2}_{2}}g'_{1} \\
                b_{1}g'_{2} & b_{2}g'_{2} & \frac{b^{2}_{1}+b^{2}_{2}}{b_{2}}g'_{2}  & -\frac{b^{2}_{1}b_{2}+b^{3}_{2}-b^{3}_{1}-b_{1}b^{2}_{2}}{b^{2}_{2}}g'_{2} \\
                c_{1}g'_{3} & c_{2}g'_{3} & \frac{c^{2}_{1}+c^{2}_{2}}{c_{2}}g'_{3} & -\frac{c^{2}_{1}c_{2}+c^{3}_{2}-c^{3}_{1}-c_{1}c^{2}_{2}}{c^{2}_{2}}g'_{3} \\
                d_{1}g'_{4} & d_{2}g'_{4} & \frac{d^{2}_{1}+d^{2}_{2}}{d_{2}}g'_{4} & -\frac{d^{2}_{1}d_{2}+d^{3}_{2}-d^{3}_{1}-d_{1}d^{2}_{2}}{d^{2}_{2}}g'_{4}
      \end{array} \right).
          \end{equation}

     We compute its determinant and we require non-vanishing of it:
     
     \begin{equation}
     \begin{gathered}
     \det{M} = -\frac{ g'_{1}g'_{2}g'_{3}g'_{4}}{a_{2}^{2}b_{2}^{2}c_{2}^{2}d_{2}^{2}} \bigg( -a_{{1}}a_{2}^{2}b_{2}^{3}c_{{2}}c_{1}^{2}d_{1}^{3}+a_{{1}}a_{2}^{2}b_{2}^{3}d_{{2}}c_{1}^{3}d_{1}^{2} \\
-a_{{1}}a_{2}^{2}c_{2}^{3}d_{{2}}d_{1}^{2}b_{1}^{3}+a_{{1}}a_{2}^{2}c_{2}^{3}b_{{2}}b_{1}^{2}d_{1}^{3}
-a_{{1}}a_{2}^{2}d_{2}^{3}b_{{2}}b_{1}^{2}c_{1}^{3}+ \\
a_{{1}}a_{2}^{2}d_{2}^{3}c_{{2}}c_{1}^{2}b_{1}^{3}
+b_{{1}}b_{2}^{2}a_{2}^{3}c_{{2}}c_{1}^{2}d_{1}^{3}-b_{{1}}b_{2}^{2}a_{2}^{3}d_{{2}}c_{1}^{3}d_{1}^{2} \\
+b_{{1}}b_{2}^{2}c_{2}^{3}d_{{2}}d_{1}^{2}a_{1}^{3}-b_{{1}}b_{2}^{2}c_{2}^{3}a_{{2}}a_{1}^{2}d_{1}^{3}
+b_{{1}}b_{2}^{2}d_{2}^{3}a_{{2}}a_{1}^{2}c_{1}^{3}- \\
b_{{1}}b_{2}^{2}d_{2}^{3}c_{{2}}c_{1}^{2}a_{1}^{3}
+c_{{1}}c_{2}^{2}a_{2}^{3}d_{{2}}d_{1}^{2}b_{1}^{3}-c_{{1}}c_{2}^{2}a_{2}^{3}b_{{2}}b_{1}^{2}d_{1}^{3} \\
-c_{{1}}c_{2}^{2}b_{2}^{3}d_{{2}}d_{1}^{2}a_{1}^{3}+c_{{1}}c_{2}^{2}b_{2}^{3}a_{{2}}a_{1}^{2}d_{1}^{3}
-c_{{1}}c_{2}^{2}d_{2}^{3}a_{{2}}a_{1}^{2}b_{1}^{3}+\\
c_{{1}}c_{2}^{2}d_{2}^{3}b_{{2}}b_{1}^{2}a_{1}^{3}
+d_{{1}}d_{2}^{2}a_{2}^{3}b_{{2}}b_{1}^{2}c_{1}^{3}-d_{{1}}d_{2}^{2}a_{2}^{3}c_{{2}}c_{1}^{2}b_{1}^{3} \\
-d_{{1}}d_{2}^{2}b_{2}^{3}a_{{2}}a_{1}^{2}c_{1}^{3}+d_{{1}}d_{2}^{2}b_{2}^{3}c_{{2}}c_{1}^{2}a_{1}^{3}
+d_{{1}}d_{2}^{2}c_{2}^{3}a_{{2}}a_{1}^{2}b_{1}^{3}-\\
d_{{1}}d_{2}^{2}c_{2}^{3}b_{{2}}b_{1}^{2}a_{1}^{3} 
 \bigg) \neq 0.  \label{wyzn_mixed}
   \end{gathered}
   \end{equation}

    As we see, it is nonzero, if $g'_{1}g'_{2}g'_{3}g'_{4} \neq 0$ and the polynomial present in (\ref{wyzn_mixed}) does not possess zeroes.
     So, if additionally: $a_{2}b_{2}c_{2}d_{2} \neq 0$, the ansatz (\ref{rozw_suma}) and the relations between the coefficients (\ref{wsp_mixed}), give some class of non-invariant solutions of mixed heavenly equation. 
     
     Analogically to the case of second heavenly equation, we extend the ansatz (\ref{rozw_suma}): 
     \begin{gather} 
       w(\eta,\xi,q,y)=\sum^{n}_{j=1} g_{j}(\Sigma_{j}), \label{rozw_suma_i_mix} 
      \end{gather}
      
      where $g_{j}$ are {\em arbitrary} functions of their arguments and: 
       
      \begin{gather}
     \Sigma_{j} = \alpha_{j} \eta + \gamma_{j} \xi + \zeta_{j} q + \lambda_{j} y + \beta_{j}. 
     \label{argumenty_i_mix} 
       \end{gather}
      
     Obviously, now the notation of the coefficients changes. We give here this change for $j=1,...,4$:
     
     \begin{equation}
     \begin{gathered}
     a_{1}=\alpha_{1}, \hspace{0.1 in} a_{2}=\gamma_{1}, \hspace{0.1 in} a_{3}=\zeta_{1}, \hspace{0.1 in} 
     a_{4} = \lambda_{1}, \\
     b_{1}=\alpha_{2}, \hspace{0.1 in} b_{2}=\gamma_{2}, \hspace{0.1 in} b_{3}=\zeta_{2}, \hspace{0.1 in} 
     b_{4} = \lambda_{2}, \\
     c_{1}=\alpha_{3}, \hspace{0.1 in} c_{2}=\gamma_{3}, \hspace{0.1 in} c_{3}=\zeta_{3}, \hspace{0.1 in} 
     c_{4} = \lambda_{3}, \\
     d_{1}=\alpha_{4}, \hspace{0.1 in} d_{2}=\gamma_{4}, \hspace{0.1 in} d_{3}=\zeta_{4}, \hspace{0.1 in} 
     d_{4} = \lambda_{4}. \label{zmiana_oznaczen_mix} 
     \end{gathered}
     \end{equation}  
     
     \vspace{0.2 in} 
     
     The ansatz (\ref{rozw_suma_i_mix}) presents some class of solutions of the system (\ref{heav_lin1})-(\ref{heav_lin3}) and consequently of (\ref{leg_mixed}), when the following relations between coefficients are satisfied:
     
     \vspace{0.2 in} 
     
     \begin{equation}
   \zeta_{j} = \frac{\alpha^{2}_{j}+\gamma^{2}_{j}}{\gamma_{j}}, \hspace{0.1 in} 
   \lambda_{j}=-\frac{\alpha^{2}_{j} \gamma_{j} + 
   \gamma^{3}_{j}-\alpha^{3}_{j} - \alpha_{j} \gamma^{2}_{j} }{\gamma^{2}_{j}} \label{wsp_i_mix}
     \end{equation}
     
     and $\theta = 1, \eta=p+t, \xi=p-t$, ($\beta_{j}$ are arbitrary constants). 
     
     Of  course, analogically, as in the case of second heavenly equation, some properties of functional series need to be satisfied,  
		 \cite{korn}. 
     Thus, (\ref{rozw_suma_i_mix}) needs to be uniformly convergent. Also here, from the requirement of 
     differentiability of (\ref{rozw_suma_i_mix}), we see that the corresponding series:
        
      \begin{equation}
      \begin{gathered}
      \sum^{n}_{j=1} \frac{\partial}{\partial \eta} g_{j} \hspace{0.08 in}, \hspace{0.05 in}
      ..., \hspace{0.08 in}  
      \sum^{n}_{j=1} \frac{\partial}{\partial y} g_{j}, \label{rozn_szereg_mix}
      \end{gathered}
      \end{equation}
      
      need to be uniformly convergent.
      
      Further, from the requirement of differentiability of the series (\ref{rozn_szereg_mix}), the  
      consecutive series, including the terms obtained by computing the derivatives in (\ref{rozn_szereg_mix})
      (for the coefficients satisfying the relations (\ref{wsp_i_mix})):
      
      \begin{gather} 
      \sum^{n}_{j=1} \frac{\partial}{\partial \eta} 
      \bigg(g'_{j} \alpha_{j}\bigg) \hspace{0.08 in}, \hspace{0.05 in} 
      \sum^{n}_{j=1} \bigg(\frac{\partial}{\partial q} g'_{j} \alpha_{j}\bigg), 
      \hspace{0.08 in} etc. \hspace{0.08 in} etc.,
      \end{gather}
      
      need also to be uniformly convergent. 
      One can check that the system (\ref{heav_lin1})-(\ref{heav_lin3}), is satisfied by 
      (\ref{rozw_suma_i_mix}), when the relations (\ref{wsp_i_mix}) hold.

     One can check also that (\ref{rozw_suma_i_mix}), for the relations 
     (\ref{wsp_i_mix}),  satisfies the mixed-heavenly equation (\ref{leg_mixed}).

     However, from the other hand, it turns out that the condition (\ref{warunek_mix}) is satisfied for the class of exact solutions given by (\ref{rozw_suma_i_mix}) and (\ref{wsp_i_mix}), when the following relations hold:
     
     \begin{gather}
     \bigg(\sum^{n}_{j=1} g''_{j} (\alpha_{j} + \gamma_{j})^{2} \bigg) \bigg(\sum^{n}_{j=1} 
     \frac{g''_{j} (\alpha^{2}_{j} + \gamma^{2}_{j})^{2}}{\gamma^{2}_{j}} \bigg) - \\  
     \bigg(\sum^{n}_{j=1} \frac{g''_{j} (\alpha^{2}_{j} + \gamma^{2}_{j})(\alpha_{j} 
     + \gamma_{j})}{\gamma_{j}} \bigg)^{2} \neq 0. \label{war_leg_i_mix}
     \end{gather}

     The condition (\ref{wyzn_mixed}) holds also for (\ref{rozw_suma_i_mix}) - (\ref{argumenty_i_mix}) and (\ref{wsp_i_mix}), 
     but of course, the notation for the coefficients $a_{j}, b_{j}, c_{j}, d_{j}, (j=1,...,4)$, changes according to 
      (\ref{zmiana_oznaczen_mix}). \\
      
     Hence, the ansatz (\ref{rozw_suma_i_mix}) with (\ref{argumenty_i_mix}), when $n$ is an arbitrary natural number,  
     and the relations between the parameters are given by (\ref{wsp_i_mix}), gives the class of the exact solutions 
	  	of (\ref{heav_lin1})-(\ref{heav_lin3}) and consequently of 
     (\ref{leg_mixed}) and these solutions depend on four variables, if $n \geq 4$, and the conditions:
     (\ref{wyzn_mixed}) (after taking into consideration the relations (\ref{zmiana_oznaczen_mix})),   
     $\gamma_{1}\gamma_{2}\gamma_{3}\gamma_{4} \neq 0$ and (\ref{war_leg_i_mix}), are satisfied. Thus, these 
     solutions are non-invariant.

		One can check by using Maple Waterloo Software that  
				the ansatz $\vartheta=\beta_{1} + \sum^{n}_{j=1} g_{j}(\Sigma_{j})$, (where $g_{j}$ are arbitrary 
				holomorphic functions of $\Sigma_{j} = a^{(n)}_{j} x^{\mu}  + \beta_{j}$), satisfies the system (\ref{heav_lin1})-(\ref{heav_lin3}) and the Legendre transformed mixed heavenly equation (\ref{leg_mixed})
				and also of , when the coefficients satisfy the relations (\ref{wsp_i_mix})
    One can also check that the condition (\ref{warunek_mix}) is satisfied for this above ansatz
		and relations (\ref{wsp_i_mix}). 
		
      \begin{thm}
	   The Legendre transformed mixed heavenly equation (\ref{leg_mixed}) (obtained by a Legendre transformation of (\ref{mixed})), 
		 possess the class  of solutions of the form of the functional series (when $\theta = 1$): 
				
		\begin{gather}
                u(x^{\mu})=\beta_{1} + \sum^{n}_{j = 1} g_{j}(\Sigma_{j}), \label{rozw_suma_niesk_mixed} 
                 \end{gather}
 
         where $u=w$ , $g_{j}$ are some arbitrary holomorphic functions of the arguments 
				$\Sigma_{m} = a^{(m)}_{\nu} x^{\nu} + \beta_{m+1}$,
				 ($n$ can be any natural number, so the series $\sum^{n}_{j = 1} g_{j}(\Sigma_{j})$ can be finite or infinite), 
				$x^{\nu}$ are proper independent variables, the series $\sum^{n}_{j = 1} g_{j}(\Sigma_{j})$ and its corresponding
				derivatives, are uniformly convergent, and $a^{(m)}_{\nu}$ are some constants 
				satisfying some relations following from satisfying of the system of algebraic equations, following from applying of the 
				decomposition method to the Legendre transformed mixed heavenly equation (\ref{leg_mixed}) .
				\end{thm}
				
				\begin{proof}
				 Analogical, as in the case of the second heavenly equation.
				\end{proof} 
   
   \subsection{Classes of exact solutions of asymmetric heavenly equation and evolution form of 
   second heavenly equation}
   
   Now we will solve the asymmetric heavenly equation (\ref{asymm}), by applying decomposition method.
   
   We insert also the ansatz (\ref{rozw_suma}), but now: $x^{1} = x, x^{2} = y, x^{3} = z, x^{4} = t$, and 
   $g_{j} \in \mathcal{C}^{2}$ ($j = 1, 2, 3, 4$), are {\em arbitrary} functions of their arguments.
   This ansatz gives the class of non-invariant solutions of (\ref{asymm}), when:
   
   \begin{equation}
   \begin{gathered}
   C=-\frac{c_{3}(Bc_{1}+Ac_{4})}{c^{2}_{1}},\\ 
	 a_{2}=\frac{a_{1}c_{2}(-Ba_{3}c^{2}_{1}+a_{1}c_{1}c_{3}B+a_{1}c_{3}c_{4}A)}{Aa_{3}c^{2}_{1}c_{4}}, \\ 
   a_{4}=\frac{a_{1}(-Ba_{3}c^{2}_{1}+a_{1}c_{1}c_{3}B+a_{1}c_{3}c_{4}A)}{Aa_{3}c^{2}_{1}}, b_{1}=0, b_{4}=0, \label{wsp_asymm}  \\
   d_{1}=0, d_{4}=0,   
    \end{gathered}
    \end{equation} 
   
   and $\beta_{k}, (k=1,...,5)$ are arbitrary constants.
   
   The Jacobian matrix has the form:
   
   \begin{equation}
          M = \left( \begin{array}{cccc}
        a_{1}g'_{1} & N_{1}g'_{1} & a_{3}g'_{1} & N_{2}g'_{1} \\
          0 & b_{2} g'_{2} & b_{3}g'_{2} &  0  \\
      c_{1}g'_{3} & c_{2}g'_{3} & c_{3}g'_{3} & c_{4}g'_{3} \\
       0 & d_{2}g'_{4} & d_{3}g'_{4} & 0
      \end{array} \right),
          \end{equation}
          
          where $N_{1}=\frac{a_{1}c_{2}}{Aa_{3}c^{2}_{1}c_{4}}(-Ba_{3}c^{2}_{1}+a_{1}c_{1}c_{3}B+Aa_{1}c_{3}c_{4}), 
          N_{2}=\frac{a_{1}}{Aa_{3}c^{2}_{1}}(-Ba_{3}c^{2}_{1}+a_{1}c_{1}c_{3}B+Aa_{1}c_{3}c_{4})$.\\

   We require non-vanishing of determinant of this above Jacobian matrix:        
   
   \begin{eqnarray}
   \begin{gathered}
   \det{M}=\frac{g'_{1}g'_{2}g'_{3}g'_{4}}{c_{1}Aa_{3}} \bigg( a_{1}(b_{2}d_{3}-b_{3}d_{2})(-c_{4} A a_{3} c_{1} - \\
	   B a_{3} c^{2}_{1} + a_{1} c_{1} c_{3} B + a_{1} c_{3} c_{4} A) \bigg) \neq 0. \label{wyzn_asymm}
  \end{gathered}
  \end{eqnarray}
   
   This above determinant is non-zero, when $g'_{1}g'_{2}g'_{3}g'_{4} \neq 0$ and the polynomial, present in 
   (\ref{wyzn_asymm}), does not possess zeroes.\\
    In the case $B=0$ we obtain the class of solutions of the evolution form of second heavenly 
    equation. Let us notice that also in this case: $B=0$, the ansatz (\ref{rozw_suma}) with the set of 
    modified relations (\ref{wsp_asymm}) (after putting $B=0$), remains the class of non-invariant 
    solutions.\\      
    Hence, the ansatz (\ref{rozw_suma}) and the relations (\ref{wsp_asymm}), give the class of exact solutions of assymetric heavenly equation (if $B \neq 0$), and of the evolution form of the second heavenly equation (if $B=0$) and these solutions depend on four variables (when (\ref{wyzn_asymm}) and $Ac_{1}a_{3}c_{4} \neq 0$ hold). Hence, these solutions are non-invariant.
    
      \subsection{Classes of exact solutions of general heavenly equation and of real section of this equation}

   \subsubsection{The case of general heavenly equation}

        We insert the ansatz (\ref{rozw_suma}) into (\ref{genHeav}), but now: $x^{1} = z^{1}, x^{2} = z^{2}, x^{3} = z^{3}, x^{4} = z^{4}$, and we wish 
       $g_{j} \in \mathcal{C}^{2}$ ($j = 1, 2, 3, 4$), are as {\em arbitrary} functions of their arguments, as it is possible. 

       The decomposition method gives a chance of obtaining several classes of exact solutions - they have the form of (\ref{rozw_suma}), where $a_{i}, b_{i}, c_{i}, d_{i} (i = 1,...,4)$ are the solutions of corresponding system of algebraic equations.\\

       I class:
       \begin{equation}
       \begin{gathered}
       a_{3} = \frac{a_{2}b_{3}c_{3}(b_{2}c_{4}-b_{4}c_{2})(\beta+\gamma)}{b_{2}c_{2}\beta(b_{3}c_{4} - 
       b_{4}c_{3})}, \\
       a_{4} = - \frac{a_{2}b_{4}c_{4}(b_{2}c_{3}-b_{3}c_{2})(\beta+\gamma)}{b_{2}c_{2}\gamma(b_{3}c_{4} - 
       b_{4}c_{3})},\\
       b_{1} = 0, c_{1} = 0, d_{1}=0, d_{2} = 0 , d_{4} = 0,\\ 
			 \beta \neq 0, \gamma \neq 0
      \end{gathered}
      \end{equation}
       
      II class:
        \begin{equation}
			  \begin{gathered}
       a_{3} = \frac{c_{3}(a_{2}c_{4}\beta + (a_{2}c_{4}-a_{4}c_{2})\gamma)}{\beta c_{2}c_{4}}, \  b_{1} = 0, \ b_{2} = 0, \\ 
      \  b_{3} = 0, \ c_{1} = 0, \ c_{3} = c_{2}, \ d_{1}=0, \ d_{3} = 0, \ d_{4} = 0.
			    \end{gathered}
       \end{equation}

     We have checked that the condition (\ref{warunek_m1}) is satisfied for these both classes.

       The conditions of non-invariance of these classes are, correspondingly

       \begin{eqnarray}
       \det{(M)} = -a_{1}d_{3}(b_{2}c_{4} -  b_{4}c_{2})g'_{1}g'_{2}g'_{3}g'_{4} \neq 0,\\
       \det{(M)} = -a_{1}b_{4}c_{2}d_{2} g'_{1} g'_{2} g'_{3} g'_{4} \neq 0.
      \end{eqnarray}

       Next, the classes of exact solutions of real general heavenly equation (\ref{genHeavReal}) are given by (\ref{rozw_suma}), however, in this case: $g_{1} = \bar{g}_{2}, g_{k} \in \mathbb{C} 
    (k=1,2), g_{n} \in \mathbb{R} (n = 3,4), c_{1}=\bar{c}_{2}, c_{3}=\bar{c}_{4}, d_{1}=\bar{d}_{2}, d_{3}=\bar{d}_{4}$, and the solutions of corresponding  system of algebraic equations:\\

       I class:

       \begin{equation}
       \begin{aligned}
       a_{2} = 0, b_{1} = 0, c_{1} = 0, c_{2} = 0, d_{1} = 0, d_{2} = 0,\\
       {\rm and} \  \beta = -\frac{\gamma b_{3} \bar{b}_{3}}{b_{4} \bar{b}_{4}},\\
        g_{1}=\bar{g}_{2}, g_{k} \in \mathbb{C} \ \  (k=1,2), \ \  g_{n} \in \mathbb{R} \ \  (n = 3,4)
       \end{aligned}
			 \end{equation} 

        II class 

        \begin{equation}
        \begin{gathered}
       a_{1}=0, a_{2}=0, a_{3} = 0,  b_{1} = 0, b_{2} = 0, b_{4} = 0\\
       {\rm and} \\   
       \beta = -\frac{\gamma(\bar{c}_{2}c_{2} \bar{d}_{4} d_{4} - c_{4} \bar{c}_{2} d_{2} \bar{d}_{4} - 
       c_{2}\bar{c}_{4} d_{4} \bar{d}_{2} + \bar{c}_{4} c_{4} d_{2} \bar{d}_{2})}{\bar{c}_{2}c_{2}\bar{d}_{4} \bar{d}_{4} 
       - \bar{c}_{2} \bar{c}_{4} d_{2} d_{4} - c_{2} c_{4} \bar{d}_{2} \bar{d}_{4} + \bar{d}_{2} \bar{c}_{4} c_{4} d_{2}}, \\
        g_{1}=\bar{g}_{2},  g_{k} \in \mathbb{C} \ \  (k = 1,2), \ \  g_{n} \in \mathbb{R} \ \ (n=3,4).
       \end{gathered}
       \end{equation}

       Obviously, we require the condition (\ref{warunek_m2}) was satisfied. It holds, when at least, some derivatives of $g_{i}, i=1,2,3,4$ are non-zero and at least one of the polynomials appearing in the numerators of the fractions, included in the  determinant of Jacobian, 
       does not possess zeroes:

       \begin{gather}
       m_{2} = b_{2} \bar{b}_{2} (b_{4} \bar{b}_{4} - b_{3} \bar{b}_{3}) g''_{1}g''_{3}  \neq 0 \ \ ({\rm{I \ class}}), \\
       m_{2} = (\bar{c}_{4} d_{4} - c_{4} \bar{d}_{4}) (\bar{c}_{2} d_{2} - c_{2} \bar{d}_{2}) g''_{3} g''_{4} \neq 0  ({\rm{II \ class}}).
      \end{gather}

       The conditions of non-invariance of these classes of solutions are correspondingly

      \begin{gather}
      \det{M} = -b_{2}\bar{b}_{2}(\bar{c}_{3}d_{3} - c_{3}\bar{d}_{3}) g'_{1}g'_{2}g'_{3}g'_{4} \neq 0, \ \  ({\rm{I \ class}})  \\
\det{M} = - b_{3} \bar{b}_{3}(c_{2} \bar{d}_{2} - \bar{c}_{2} d_{2}) g'_{1}g'_{2}g'_{3}g'_{4} \neq 0 \hspace{0.1 in}  \ ({\rm{\ II \ class}}). 
      \end{gather}

  We stress here that we have obtained these above classes of solutions of (\ref{genHeav}), i.e. functionally invariant solutions of this equation, without imposing the differential constraint (\ref{constraint}).

       \subsubsection{The case of one of the real sections of general heavenly equation}
   
        We insert the ansatz (\ref{rozw_suma}) into (\ref{genHeavRealSec2}), but now: $x^{1} = z^{1}, x^{2} = \bar{z}^{1}, x^{3} = z^{3}, x^{4} = \bar{z}^{2}, \omega \in 
        \mathbb{R}$, and, as previously, we wish  $g_{j} \in \mathcal{C}^{2}$ ($j = 1, 2, 3, 4$), are as {\em arbitrary} functions of their arguments, as it is possible. 
        It turns out that certain non-invariant class of the solutions for (\ref{genHeavRealSec2}), for $m \neq 0$, is given by the ansatzes (\ref{rozw_suma}), when respectively: 

       \begin{equation}
       \begin{gathered}
        g_{1} = \bar{g}_{2}, g_{3}=\bar{g}_{4}, g_{n} \in \mathbb{C} \ \ (n=1,2,3,4), \\ 
		  	\delta = \sqrt{\frac{b_{4} \bar{b}_{4}}{a_{4}  
        \bar{a}_{4}}}, a_{1} = 0, 
        a_{3} = \bar{b}_{4},  
        b_{1}=\bar{a}_{2}, b_{2} = 0,\\ 
        b_{3} = \bar{a}_{4}, c_{1} = 0, c_{2} = 0, c_{3} = \bar{d}_{4}, c_{4} = \bar{d}_{3}, d_{1} = 0, \\
				d_{2} = 0, \beta_{k} \in 
       \mathbb{C} \ \ (k=1,2,3,4),\\ 
       \beta_{1} = \bar{\beta}_{2}, \beta_{3} = \bar{\beta}_{4}.
       \end{gathered}
       \end{equation}
   
      The condition of non-invariance of these solutions, is:

     \begin{equation}
     \det{(M)} =  a_{2} \bar{a}_{2} (d_{3}\bar{d}_{3} - d_{4} \bar{d}_{4})  g'_{1} g'_{2} g'_{3} g'_{4} \neq 0.
     \end{equation}

      The condition (\ref{warunek_m2})  is satisfied, when

    \begin{equation}
     m_{2} = a_{2} \bar{a}_{2}(a_{4} \bar{a}_{4} - b_{4} \bar{b}_{4})g''_{1} g''_{2} \neq 0.
    \end{equation}

    \subsection{The criterion for non-invariance of found solutions}
    
      From the results obtained in previous sections of this paper, especially from the forms of the Jacobians 
      for the found classes of exact solutions, the criterion for non-invariance of the 
      solutions, belonging to the above mentioned classes, follows immediately:

      \begin{coro}      
      
      Let the ansatz: 
      
      \begin{gather} 
        u(x^{k})=\sum^{n}_{j=1} g_{j}(\Sigma_{j}), \hspace{0.1 in} k=1,2,3,4, 
       \label{rozw_suma_i2} 
      \end{gather}
      
			 where: $\Sigma_{j} = \alpha_{j} x^{1} + \gamma_{j} x^{2} + \zeta_{j} x^{3} + \lambda_{j} x^{4} + 
      \beta_{j}$, the coefficients $\alpha_{j}, \gamma_{j}, \zeta_{j}, \lambda_{j}, \beta_{j}$ satisfy some relations, 
      and $n=4$ (for the equations: elliptic and hyperbolic complex Monge-Amp$\grave{e}$re, asymmetric 
      heavenly, evolution form of heavenly equation, general heavenly equation, real general heavenly equation and one of the real sections of 
      general heavenly equation) 
      or $n\geq 4$ (for second heavenly equation and Legendre transformed mixed heavenly equation), $g_{j} \in \mathcal{C}^{2}$ are 
      {\em arbitrary} functions (however: in the case of elliptic complex Monge-Amp$\grave{e}$re equation, $g_{1},g_{2}$ are the 
      square 
      functions of their arguments and in the case of higher symmetry for the second heavenly 
      equation, the ansatz (\ref{rozw_suma_i2}) needs to be convergent series, some series including
      terms obtained after computing derivatives of (\ref{rozw_suma_i2}), need to be uniformly convergent, 
      moreover, in the case of elliptic and hyperbolic complex Monge-Amp$\grave{e}$re equation, the solutions, belonging to the 
      corresponding 
      classes, given by (\ref{rozw_suma_i2}), need to be real), 
      gives the class of exact solutions of the equations: elliptic and hyperbolic complex Monge-Amp$\grave{e}$re one, second  
      heavenly, 
      mixed heavenly, asymmetric heavenly, evolution form of second heavenly equation, general heavenly equation, real general heavenly 
      equation 
      and one of the real sections of general heavenly equation, when corresponding conditions of 
      existence of Legendre transformation are satisfied (in the case of the equations: hyperbolic complex Monge-Amp$\grave{e}$re, 
      second heavenly, mixed heavenly). \\
      These solutions are non-invariant, if first derivatives of the all functions $g_{n}, (n=1,...,4)$, are non-zero, 
      the polynomials included in Jacobians, corresponding to each of these above mentioned classes of solutions, do not possess 
      zeroes
      and the coefficients $\alpha_{j}, \gamma_{j}, \zeta_{j}, \lambda_{j}$ satisfy some additional relations (in the case of the equations:  
      second heavenly and mixed, also uniform convergence of the series and their proper derivatives, is required).

      \end{coro}

    \section{Conclusions}
    
      We applied decomposition method for finding of classes of exact solutions (functionally invariant solutions) of heavenly equations:   second heavenly, mixed heavenly, asymmetric heavenly, evolution form of second heavenly equation, general heavenly equation, real  general heavenly equation and one of the real sections of general heavenly equation. For each of these equations, we have obtained the algebraic determining system, following from inserting the ansatz into investigated equation. 
      Apart from satisfying of such algebraic determining system and the condition of non-invariance of wanted solutions, belonging to our classes (which implicated non-vanishing of the Jacobians), in the cases of the equations: second heavenly, mixed heavenly, general heavenly equation, real  general heavenly equation and one of the real sections of general heavenly equation, some additional conditions must be satisfied by the functions included in ansatz and by the coefficients. It depends on the investigated equation:
      
      \begin{enumerate}
			\item elliptic complex Monge-Amp$\grave{e}$re equation - the 
      condition of reality of the solutions,
      \item	hyperbolic complex Monge-Amp$\grave{e}$re equation - the condition of existence of Legendre transformation and the 
      condition of reality of the solutions,
      \item second heavenly equation - the condition of existence of Legendre transformation;  
      the class of exact solutions is given by general functional series (for any $n$); obviously,  
      this series needs to be uniformly convergent, some series obtained by twice differentiating of this series, need to 
      be also uniformly convergent, 
      \item mixed-heavenly equation - the class of exact solutions is given by general functional series (for any $n$); 
			 the condition of existence of Legendre transformation,
      of course, the condition of unifom convergence is analogical to the case of the second heavenly equation,
			\item asymmetric heavenly equation - the condition (\ref{wyzn_asymm}) 
     \item general heavenly equation -  the condition (\ref{warunek_m1}) 
     \item real general heavenly equation and one of the real sections of general heavenly equation - the condition 
     (\ref{warunek_m2}) and  the condition of reality of the solutions.
      \end{enumerate}
      
			Obviously, in the case of the equations: hyperbolic complex Monge-Amp$\grave{e}$re, second heavenly equation, mixed heavenly equation, the corresponding conditions of 
      existence of Legendre transformation should also be satisfied.\\
			
			The subclasses I and II (for the case of higher symmetry) was applied to construct in \cite{stepien2017}, some exact solutions
			of self-dual Yang-Mills (SDYM) equations, in the non-$R$ gauge case, owing to the reduction
			of SDYM equations to second heavenly equation, done in \cite{PlebanskiPrzanowskiGarciaCompean}.

   We tried to keep the generality of the functions including in the ansatz, as it was possible, too.\\

      To sum up, we have found some new classes of exact, non-invariant solutions of each of all above 
      mentioned equations and we have established also the criterion for the non-invariance of the solutions,
      belonging to these classes.\\
      
      Moreover, although there are several methods of solving of nonlinear partial differential equations \cite{Kudryashow}, 
			\cite{Meleshko}, \cite{Olver}, \cite{PolyaninZaitsev}, \cite{Rubina}, and the decomposition method, applied in this paper, is not 
      general method, one can say that this method can offer sometimes more easy way of finding of exact solutions of some nonlinear 
      partial differential equations, like heavenly equations, in comparison with other exact methods. \\
      In some cases, this method can give a possibility of finding classes of exact solutions of given nonlinear PDE, by applying
      this method directly to the equation, without the necessity of linearization of this PDE. A good example
      can be here \\     
      Of course, mentioned above solutions, of the equations investigated in this paper, are not first found functionally invariant 
      solutions of these equations. Actually, some functionally invariant solutions were found to the equations:
      second heavenly in \cite{dun_mason} . 
      However, they possess different form, than the solutions presented in the section 4 of the current paper.\\
			It is easy to check that the functionally-invariant solutions found in the current paper
			for the second heavenly equation, have more general functional form than the multikink solutions found in \cite{wazwaz}.

  \appendix

  \section{Condition (\ref{warunek_mix}) for the class of solutions, given by (\ref{rozw_suma}) and 
  (\ref{wsp_mixed}), of mixed heavenly equation}
  
   \begin{gather}
     \frac{1}{a_{2}^{2}b_{2}^{2}c_{2}^{2}d_{2}^{2}}(N_{1} g''_{1}g''_{2}+N_{2} g''_{1} g''_{3}
+N_{3} g''_{1}g''_{4}+N_{4} g''_{2}g''_{3}+N_{5} g''_{2}g''_{4}+N_{6} g''_{3}g''_{4}) \neq 0, \label{war_leg_mix}
     \end{gather}

     where $N_{1}= b_{2}^{4}c_{2}^{2}d_{2}^{2}a_{1}^{4}+a_{2}^{4}c_{2}^{2}d_{2}^{2}b_{1}^{4}
+a_{1}^{2}a_{2}^{2}c_{2}^{2}d_{2}^{2}b_{1}^{4}+2\,a_{{1}}a_{2}^{3}c_{2}^{2}d_{2}^{2}b_{1}^{4}
+a_{2}^{4}c_{2}^{2}d_{2}^{2}b_{1}^{2}b_{2}^{2}+a_{1}^{2}a_{2}^{2}c_{2}^{2}d_{2}^{2}b_{2}^{4}
-2b_{{2}}c_{2}^{2}d_{2}^{2}a_{{1}}a_{2}^{3}b_{1}^{3}-2b_{2}^{3}c_{2}^{2}d_{2}^{2}a_{{1}}a_{2}^{3}b_{{1}}
+2b_{{1}}b_{2}^{3}c_{2}^{2}d_{2}^{2}a_{1}^{2}a_{2}^{2}-2b_{{2}}c_{2}^{2}d_{2}^{2}a_{1}^{3}a_{{2}}b_{1}^{3}
-2b_{2}^{2}c_{2}^{2}d_{2}^{2}a_{1}^{3}a_{{2}}b_{1}^{2}-2b_{2}^{3}c_{2}^{2}d_{2}^{2}a_{1}^{3}a_{{2}}b_{{1}}
+2a_{1}^{2}a_{2}^{2}c_{2}^{2}d_{2}^{2}b_{1}^{2}b_{2}^{2}+2a_{{1}}a_{2}^{3}c_{2}^{2}d_{2}^{2}b_{1}^{2}b_{2}^{2}
+b_{1}^{2}b_{2}^{2}c_{2}^{2}d_{2}^{2}a_{1}^{4}+2b_{{1}}b_{2}^{3}c_{2}^{2}d_{2}^{2}a_{1}^{4}
-2b_{2}^{4}c_{2}^{2}d_{2}^{2}a_{1}^{3}a_{{2}}-2b_{{2}}c_{2}^{2}d_{2}^{2}a_{2}^{4}b_{1}^{3}
-2b_{{2}}c_{2}^{2}d_{2}^{2}a_{1}^{2}a_{2}^{2}b_{1}^{3}$,\\ 
$N_{2}= a_{1}^{2}a_{2}^{2}b_{2}^{2}d_{2}^{2}c_{2}^{4}+2a_{{1}}a_{2}^{3}b_{2}^{2}d_{2}^{2}c_{1}^{4}
+a_{2}^{4}b_{2}^{2}d_{2}^{2}c_{1}^{4}+c_{2}^{4}b_{2}^{2}d_{2}^{2}a_{1}^{4}
+2c_{{1}}c_{2}^{3}b_{2}^{2}d_{2}^{2}a_{1}^{4}-2b_{2}^{2}c_{2}^{4}d_{2}^{2}a_{1}^{3}a_{{2}}
+a_{2}^{4}b_{2}^{2}d_{2}^{2}c_{1}^{2}c_{2}^{2}+c_{1}^{2}b_{2}^{2}c_{2}^{2}d_{2}^{2}a_{1}^{4}
+a_{1}^{2}a_{2}^{2}b_{2}^{2}d_{2}^{2}c_{1}^{4}-2b_{2}^{2}c_{{2}}d_{2}^{2}a_{1}^{2}a_{2}^{2}c_{1}^{3}
-2b_{2}^{2}c_{{2}}d_{2}^{2}a_{2}^{4}c_{1}^{3}+2a_{{1}}a_{2}^{3}b_{2}^{2}d_{2}^{2}c_{1}^{2}c_{2}^{2}
+2c_{{1}}c_{2}^{3}b_{2}^{2}d_{2}^{2}a_{1}^{2}a_{2}^{2}-2b_{2}^{2}c_{{2}}d_{2}^{2}a_{1}^{3}a_{{2}}c_{1}^{3}
-2b_{2}^{2}c_{{2}}d_{2}^{2}a_{{1}}a_{2}^{3}c_{1}^{3}-2b_{2}^{2}c_{2}^{3}d_{2}^{2}a_{{1}}a_{2}^{3}c_{{1}}
+2a_{1}^{2}a_{2}^{2}b_{2}^{2}d_{2}^{2}c_{1}^{2}c_{2}^{2}-2b_{2}^{2}c_{2}^{2}d_{2}^{2}a_{1}^{3}a_{{2}}c_{1}^{2}
-2b_{2}^{2}c_{2}^{3}d_{2}^{2}a_{1}^{3}a_{{2}}c_{{1}}$,\\ 
$N_{3}=d_{1}^{2}b_{2}^{2}c_{2}^{2}d_{2}^{2}a_{1}^{4}+2d_{{1}}d_{2}^{3}b_{2}^{2}c_{2}^{2}a_{1}^{4}
+2a_{{1}}a_{2}^{3}b_{2}^{2}c_{2}^{2}d_{1}^{2}d_{2}^{2}+d_{2}^{4}b_{2}^{2}c_{2}^{2}a_{1}^{4}
+a_{1}^{2}a_{2}^{2}b_{2}^{2}c_{2}^{2}d_{2}^{4}+2d_{{1}}d_{2}^{3}b_{2}^{2}c_{2}^{2}a_{1}^{2}a_{2}^{2}
-2b_{2}^{2}c_{2}^{2}d_{2}^{4}a_{1}^{3}a_{{2}}-2b_{2}^{2}c_{2}^{2}d_{{2}}a_{2}^{4}d_{1}^{3}
+a_{2}^{4}b_{2}^{2}c_{2}^{2}d_{1}^{4}+a_{2}^{4}b_{2}^{2}c_{2}^{2}d_{1}^{2}d_{2}^{2}
+2a_{1}^{2}a_{2}^{2}b_{2}^{2}c_{2}^{2}d_{1}^{2}d_{2}^{2}-2b_{2}^{2}c_{2}^{2}d_{2}^{3}a_{1}^{3}a_{{2}}d_{{1}}
-2b_{2}^{2}c_{2}^{2}d_{{2}}a_{1}^{2}a_{2}^{2}d_{1}^{3}-2b_{2}^{2}c_{2}^{2}d_{{2}}a_{{1}}a_{2}^{3}d_{1}^{3}
-2b_{2}^{2}c_{2}^{2}d_{2}^{2}a_{1}^{3}a_{{2}}d_{1}^{2}-2b_{2}^{2}c_{2}^{2}d_{{2}}a_{1}^{3}a_{{2}}d_{1}^{3}
+a_{1}^{2}a_{2}^{2}b_{2}^{2}c_{2}^{2}d_{1}^{4}-2b_{2}^{2}c_{2}^{2}d_{2}^{3}a_{{1}}a_{2}^{3}d_{{1}}
+2a_{{1}}a_{2}^{3}b_{2}^{2}c_{2}^{2}d_{1}^{4}$,\\ 
$N_{4}=2b_{{1}}b_{2}^{3}a_{2}^{2}d_{2}^{2}c_{1}^{4}+b_{1}^{2}a_{2}^{2}b_{2}^{2}d_{2}^{2}c_{2}^{4}
+b_{1}^{2}a_{2}^{2}b_{2}^{2}d_{2}^{2}c_{1}^{4}-2a_{2}^{2}c_{2}^{3}d_{2}^{2}b_{1}^{3}b_{{2}}c_{{1}}
-2a_{2}^{2}c_{{2}}d_{2}^{2}b_{1}^{2}b_{2}^{2}c_{1}^{3}-2a_{2}^{2}c_{{2}}d_{2}^{2}b_{2}^{4}c_{1}^{3}
-2a_{2}^{2}c_{2}^{4}d_{2}^{2}b_{1}^{3}b_{{2}}-2a_{2}^{2}c_{{2}}d_{2}^{2}b_{1}^{3}b_{{2}}c_{1}^{3}
-2a_{2}^{2}c_{2}^{2}d_{2}^{2}b_{1}^{3}b_{{2}}c_{1}^{2}+2c_{{1}}c_{2}^{3}a_{2}^{2}d_{2}^{2}b_{1}^{2}b_{2}^{2}
+b_{2}^{4}a_{2}^{2}d_{2}^{2}c_{1}^{2}c_{2}^{2}+b_{2}^{4}a_{2}^{2}d_{2}^{2}c_{1}^{4}
+c_{2}^{4}a_{2}^{2}d_{2}^{2}b_{1}^{4}+2c_{{1}}c_{2}^{3}a_{2}^{2}d_{2}^{2}b_{1}^{4}
+2b_{{1}}b_{2}^{3}a_{2}^{2}d_{2}^{2}c_{1}^{2}c_{2}^{2}+2b_{1}^{2}a_{2}^{2}b_{2}^{2}d_{2}^{2}c_{1}^{2}c_{2}^{2}
+c_{1}^{2}a_{2}^{2}c_{2}^{2}d_{2}^{2}b_{1}^{4}-2a_{2}^{2}c_{{2}}d_{2}^{2}b_{{1}}b_{2}^{3}c_{1}^{3}
-2a_{2}^{2}c_{2}^{3}d_{2}^{2}b_{{1}}b_{2}^{3}c_{{1}}$,\\ 
$N_{5}=b_{1}^{2}a_{2}^{2}b_{2}^{2}c_{2}^{2}d_{2}^{4}-2a_{2}^{2}c_{2}^{2}d_{{2}}b_{2}^{4}d_{1}^{3}
+2b_{{1}}b_{2}^{3}a_{2}^{2}c_{2}^{2}d_{1}^{2}d_{2}^{2}+b_{1}^{2}a_{2}^{2}b_{2}^{2}c_{2}^{2}d_{1}^{4}
-2a_{2}^{2}c_{2}^{2}d_{2}^{4}b_{1}^{3}b_{{2}}+b_{2}^{4}a_{2}^{2}c_{2}^{2}d_{1}^{2}d_{2}^{2}
+2d_{{1}}d_{2}^{3}a_{2}^{2}c_{2}^{2}b_{1}^{2}b_{2}^{2}+2d_{{1}}d_{2}^{3}a_{2}^{2}c_{2}^{2}b_{1}^{4}
+2b_{{1}}b_{2}^{3}a_{2}^{2}c_{2}^{2}d_{1}^{4}+2b_{1}^{2}a_{2}^{2}b_{2}^{2}c_{2}^{2}d_{1}^{2}d_{2}^{2}
+d_{1}^{2}a_{2}^{2}c_{2}^{2}d_{2}^{2}b_{1}^{4}+d_{2}^{4}a_{2}^{2}c_{2}^{2}b_{1}^{4}
-2a_{2}^{2}c_{2}^{2}d_{2}^{3}b_{{1}}b_{2}^{3}d_{{1}}-2a_{2}^{2}c_{2}^{2}d_{{2}}b_{1}^{3}b_{{2}}d_{1}^{3}
+b_{2}^{4}a_{2}^{2}c_{2}^{2}d_{1}^{4}-2a_{2}^{2}c_{2}^{2}d_{2}^{3}b_{1}^{3}b_{{2}}d_{{1}}
-2a_{2}^{2}c_{2}^{2}d_{{2}}b_{1}^{2}b_{2}^{2}d_{1}^{3}-2a_{2}^{2}c_{2}^{2}d_{{2}}b_{{1}}b_{2}^{3}d_{1}^{3}
-2a_{2}^{2}c_{2}^{2}d_{2}^{2}b_{1}^{3}b_{{2}}d_{1}^{2}$,\\ 
$N_{6}=2c_{{1}}c_{2}^{3}a_{2}^{2}b_{2}^{2}d_{1}^{4}+c_{2}^{4}a_{2}^{2}b_{2}^{2}d_{1}^{2}d_{2}^{2}
+2d_{{1}}d_{2}^{3}a_{2}^{2}b_{2}^{2}c_{1}^{4}-2a_{2}^{2}b_{2}^{2}d_{2}^{3}c_{1}^{3}c_{{2}}d_{{1}}
-2a_{2}^{2}b_{2}^{2}d_{{2}}c_{1}^{2}c_{2}^{2}d_{1}^{3}+c_{1}^{2}a_{2}^{2}b_{2}^{2}c_{2}^{2}d_{2}^{4}
+d_{1}^{2}a_{2}^{2}b_{2}^{2}d_{2}^{2}c_{1}^{4}-2a_{2}^{2}b_{2}^{2}d_{2}^{2}c_{1}^{3}c_{{2}}d_{1}^{2}
-2a_{2}^{2}b_{2}^{2}d_{{2}}c_{2}^{4}d_{1}^{3}+c_{1}^{2}a_{2}^{2}b_{2}^{2}c_{2}^{2}d_{1}^{4}
-2a_{2}^{2}b_{2}^{2}d_{{2}}c_{1}^{3}c_{{2}}d_{1}^{3}-2a_{2}^{2}b_{2}^{2}d_{2}^{4}c_{1}^{3}c_{{2}}
-2a_{2}^{2}b_{2}^{2}d_{{2}}c_{{1}}c_{2}^{3}d_{1}^{3}-2a_{2}^{2}b_{2}^{2}d_{2}^{3}c_{{1}}c_{2}^{3}d_{{1}}
+c_{2}^{4}a_{2}^{2}b_{2}^{2}d_{1}^{4}+d_{2}^{4}a_{2}^{2}b_{2}^{2}c_{1}^{4}
+2d_{{1}}d_{2}^{3}a_{2}^{2}b_{2}^{2}c_{1}^{2}c_{2}^{2}+2c_{1}^{2}a_{2}^{2}b_{2}^{2}c_{2}^{2}d_{1}^{2}d_{2}^{2}
+2c_{{1}}c_{2}^{3}a_{2}^{2}b_{2}^{2}d_{1}^{2}d_{2}^{2}$.

   \vspace{0.4 in} 
  
  \section{Acknowledgements}
  
  The author thanks to Prof. M. Sheftel for very interesting discussion. 
	The author is indebted also to Dr Z. Lisowski, Prof. V. Mityushev and Prof. K. Sokalski for their valuable remarks. 


  \section{Computational resources} 
  
   The computations were carried out, by using Waterloo MAPLE Software on the computers: "mars"
   and "saturn" (No. of grants: MNiI/IBM BC HS21/AP/057/2008 and  MNiI/Sun6800/WSP/008/2005, correspondingly)
   in ACK-Cyfronet AGH in Krak\'{o}w (Poland).
	 Some part of the computations was done by using Waterloo Maple Software,
	 owing to the financial support, provided by The Pedagogical University of Cracow, within some research project (the leader of this 
	 project: Dr K. Rajchel).
   This research was supported also by PL-Grid Infrastructure.  
   The computations were carried out also in Interdisciplinary Centre for Mathematical and Computer Modelling 
   (ICM), within the grant No. G31-6.

\section*{References}



\end{document}